\documentclass[reprint,amsmath,amssymb,aps,pra,superscriptaddress]{revtex4-2}

\usepackage{chemformula} % Formula subscripts using \ch{}
\usepackage[T1]{fontenc} % Use modern font encodings
\usepackage{graphicx}
\usepackage{dcolumn}% Align table columns on decimal point
\usepackage{bm}% bold math
\usepackage[colorlinks,linkcolor=blue,citecolor=blue,urlcolor=blue]{hyperref}% add hypertext capabilities

\begin{document}

\title{Towards real-world applications of levitated optomechanics}

\author{Yuanbin Jin}
 \affiliation{Department of Physics and Astronomy, Purdue University, West Lafayette, Indiana 47907, USA.}
\author{Kunhong Shen}
 \affiliation{Department of Physics and Astronomy, Purdue University, West Lafayette, Indiana 47907, USA.}
\author{Peng Ju}
 \affiliation{Department of Physics and Astronomy, Purdue University, West Lafayette, Indiana 47907, USA.}
\author{Tongcang Li}
 \email{tcli@purdue.edu}
 \affiliation{Department of Physics and Astronomy, Purdue University, West Lafayette, Indiana 47907, USA.}
 \affiliation{Elmore Family School of Electrical and Computer Engineering, Purdue University, West Lafayette, Indiana 47907, USA.}
 \affiliation{Birck Nanotechnology Center, Purdue University, West Lafayette, Indiana 47907, USA.}
 \affiliation{Purdue Quantum Science and Engineering Institute, Purdue University, West Lafayette, Indiana 47907, USA.}

\date{\today}

\begin{abstract}
Levitated optomechanics, a rapidly expanding field that employs light to monitor and manipulate the mechanical motion of levitated objects, is increasingly relevant across physics, engineering, and other fields. This technique, which involves levitating micro- and nano-scale objects in a vacuum where they exhibit high-quality motion, provides an essential platform for precision measurements. Noted for their ultra-high sensitivity, levitated particles hold potential for a wide range of real-world applications. This perspective article briefly introduces the principle of optical levitation and the dynamics of levitated particles. It then reviews the emerging applications of levitated  particles in ultrasensitive force and torque measurements, acceleration and rotation sensing, electric and magnetic field detection, scanning probe microscopy, localized vacuum pressure gauging, acoustic transduction, and chemical and biological sensing. Moreover, we discuss the present challenges and explore opportunities to minimize and integrate levitation systems for broader applications. We also briefly review optomechanics with ion traps and magnetic traps which can levitate particles in high vacuum without laser heating. 
\end{abstract}

\maketitle

\section{Introduction}

Optical levitation of small particles in vacuum using laser radiation pressure was first experimentally demonstrated by A. Ashkin and J. M. Dziedzic in 1970's \cite{Ashkin1971Levitation}. Decades later, Li \textit{et al.} measured the instantaneous velocity of the Brownian motion of an optically levitated glass microsphere \cite{Li2010Measurement} and cooled its center-of-mass (CoM) motion to millikelvin temperatures using feedback cooling \cite{Li2011Millikelvin}.  The core of levitation systems lies in their capability to levitate particles, typically at micro- and nano-scale, and precisely monitor and control dynamics using optical forces and other techniques. The CoM motion of a levitated particle has been cooled with various methods, i.e., force feedback cooling \cite{Li2011Millikelvin,Bang2020Five,Tebbenjohanns2019Cold,Conangla2019Optimal,Magrini2021Real,Tebbenjohanns2021Quantum,Blakemore2022Librational,Liska2023Cold}, parametric feedback cooling \cite{Gieseler2012Subkelvin,Zheng2019Cooling,Gao2024Feedback,Arita2022All}, and cavity cooling \cite{Millen2015Cavity,Delic2020Cooling,Pontin2023Simultaneous,Piotrowski2023Simultaneous}. In 2020, the CoM motion temperature of a levitated particle in high vacuum was cooled to its quantum ground-state  \cite{Delic2020Cooling}, bringing levitated optomechanics to the quantum regime \cite{Magrini2021Real,Tebbenjohanns2021Quantum}. 

\begin{figure*}
	\includegraphics[width=0.8\textwidth]{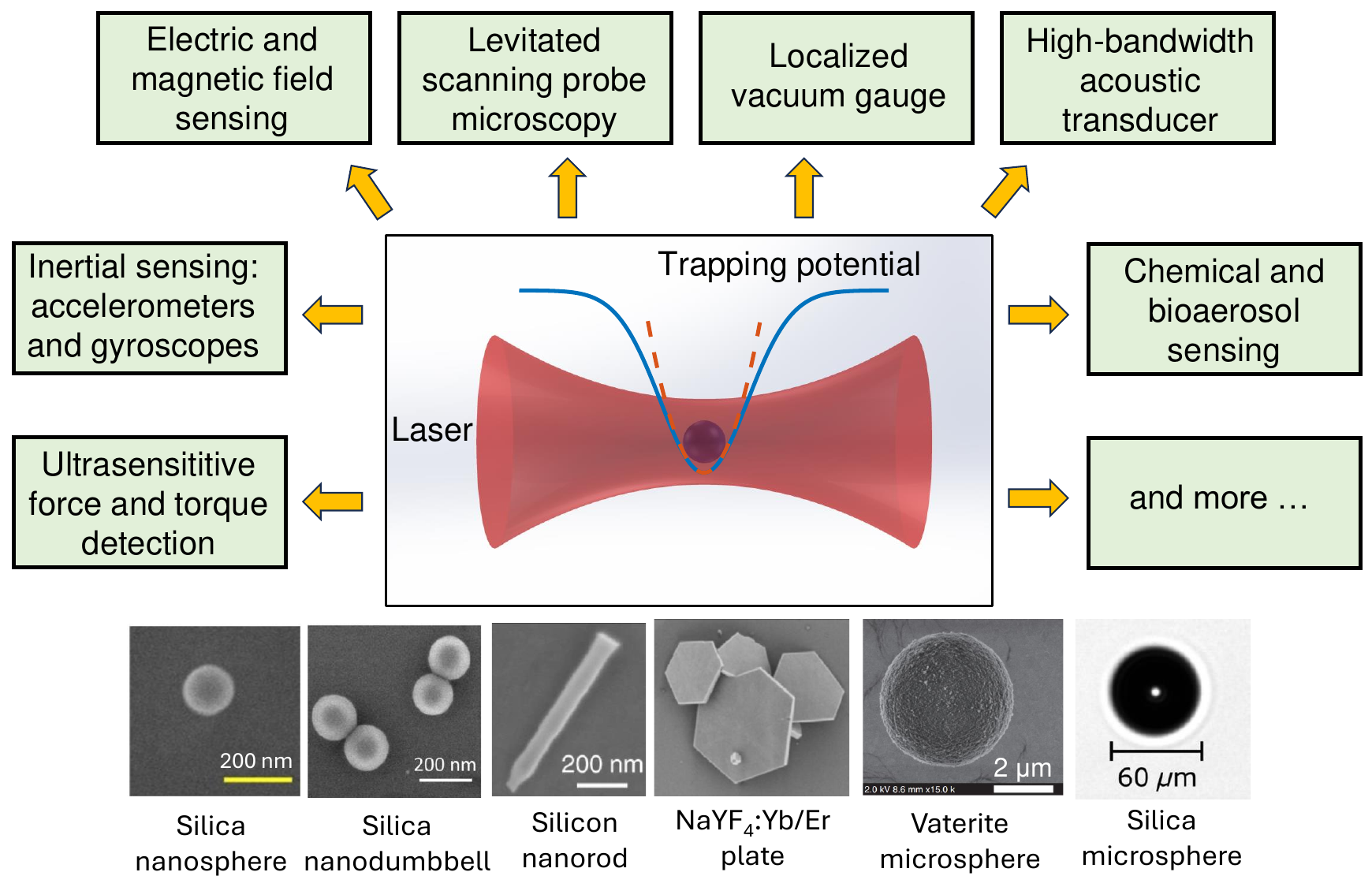}
	\caption{An overview of applications of levitated optomechanics. Levitated particles, typically at micro- and nano-scales, offer various applications in sensing and manipulation, such as force and torque measurement, inertia sensing, electric and magnetic field sensing, levitated scanning probe microscopu, vacuum gauge, chemical and bioaerosol sensing, and others. The multifaceted applications of levitation highlight its adaptability, wherein the selection of particles is crucial to the particular objectives of each application based on the distinctive properties of particles. The bottom row of this figure shows some typical scanning electron microscope (SEM) images of levitated particles: silica nanosphere \cite{Ahn2020Ultrasensitive}, silica nanodumbbell \cite{Ahn2018Optically}, silicon nanorod \cite{Kuhn2017Optically}, NaYF$_4$:Yb/Er plate \cite{Winstone2022Optical}, vaterite microsphere \cite{Arita2013Laser}, and silica microsphere \cite{Lewandowski2021high}. SEM images are reprinted with permissions from \cite{Ahn2018Optically}, Copyright (2018) by the American Physical Society;  \cite{Winstone2022Optical}, Copyright (2022) by the American Physical Society; and \cite{Lewandowski2021high}, Copyright (2021) by the American Physical Society.}
	\label{fig:overview}
\end{figure*}

Due to the absence of mechanical contact with the thermal environment, levitation systems in vacuum exhibit an ultra-high mechanical quality factor, holding promise for innovative applications. In addition, the motion of levitated particles can be optically detected at the quantum limit, making them particularly suitable for sensing applications (Fig.~\ref{fig:overview}). Levitated systems are good in measurement of weak forces and torques \cite{Ranjit2016Zeptonewton,Monteiro2020Force,Ahn2020Ultrasensitive,Ju2023Near,Hebestreit2018Sensing,Timberlake2019Acceleration,Priel2022Dipole,Zhu2023Nanoscale,Liang2023Yoctonewton}.   Recently, a force detection sensitivity of $(6.3 \pm 1.6)  \times 10^{-21}$ $\mathrm{N/\sqrt{Hz}}$ \cite{Liang2023Yoctonewton} and a torque detection  sensitivity of $\left( 4.2 \pm 1.2 \right) \times 10^{-27}$ $\mathrm{N m / \sqrt{Hz}}$ were experimentally demonstrated with optically levitated nanoparticles \cite{Ahn2020Ultrasensitive}.
Besides applications in fundamental researches, such as exploration of non-Newtonian gravity \cite{Geraci2010Short,Chen2022constraining} and probing dark mater \cite{Carney2021Mechanical,Afek2022Coherent}, levitated particles can serve as inertial sensors for detecting minute changes in acceleration and rotation \cite{Monteiro2020Force,Li2023Collective}. An optical levitation system has been packed as a small-scale accelerometer and evaluated by testing it on a vehicle running on a road \cite{Han2023Feedback}, providing an example for real-world applications. Levitated systems have also been used for electric field \cite{Zhu2023Nanoscale} and magnetic field \cite{Jiang2020Superconducting,Ahrens2024Levitated} detection, scanning probe microscopy \cite{Ju2023Near}, localized vacuum pressure gauging \cite{Blakemore2020Absolute}, acoustic transduction \cite{Hillberry2024acoustic}, and chemical and biological sensing \cite{Ai2022Characterization}.

Various trapped particles with different characteristics provide versatile options to meet specific objectives of individual applications. The bottom row in Fig.~\ref{fig:overview} lists some typical levitated particles, such as silica nanospheres \cite{Delic2020Cooling}, silica nanodumbbells \cite{Ahn2018Optically}, silicon nanorods \cite{Kuhn2017Optically}, NaYF$_4$:Yb/Er plates \cite{Winstone2022Optical}, vaterite microspheres \cite{Arita2013Laser}, and silica microspheres \cite{Lewandowski2021high}, among others. More details of application with different kinds of particle will be discussed in Sec.~\ref{sec:different_particles}. In addition, levitated particles with embedded electron spin qubits, such as diamond nitrogen-vacancy (NV) centers \cite{Neukirch2015Multi,Hoang2016Electron,Jin2024Quantum}, are good for creating matter-wave interferometers \cite{Yin2013Large,Bose2017Spin,Marletto2017Gravitationally,Ledbetter2012Gyroscopes,Zhang2023Highly,Ma2017Proposal,Rusconi2022Spin},  investigating strong spin-mechanical coupling \cite{Delord2020Spin}, and probing quantum geometric phase \cite{Chudo2014Observation,Wood2017Magnetic,Jin2024Quantum}.

There have been several review articles on levitated optomechanics \cite{yin2013optomechanics,gieseler2018levitated,millen2020optomechanics,Gonzalez2021Levitodynamics,moore2021searching,winstone2023levitated}. However, these review articles \cite{yin2013optomechanics,gieseler2018levitated,millen2020optomechanics,Gonzalez2021Levitodynamics,moore2021searching,winstone2023levitated} focused on the foundations of levitated optomechanics and its applications in fundamental research. In this perspective article, we focus on the practical applications of levitated optomechanics. We discuss the emerging applications of levitated particles in force and torque measurement, inertial sensing, electric and magnetic field detection, levitated scanning probe microscopy, vacuum gauging, acoustic transduction, and chemical and bioaerosol sensing. Furthermore, the integration and minimization of levitation systems become crucial for the practical implementation of these levitated sensors in real-world scenarios. We review the  challenges in creating compact levitation systems, and discuss various methods of particle launching and on-chip levitation, and hybrid systems with ion traps and magnetic traps to address those  challenges for real-world applications of levitated optomechanics.

\section{Optical levitation}

\begin{figure*}
	\includegraphics[width=0.8\textwidth]{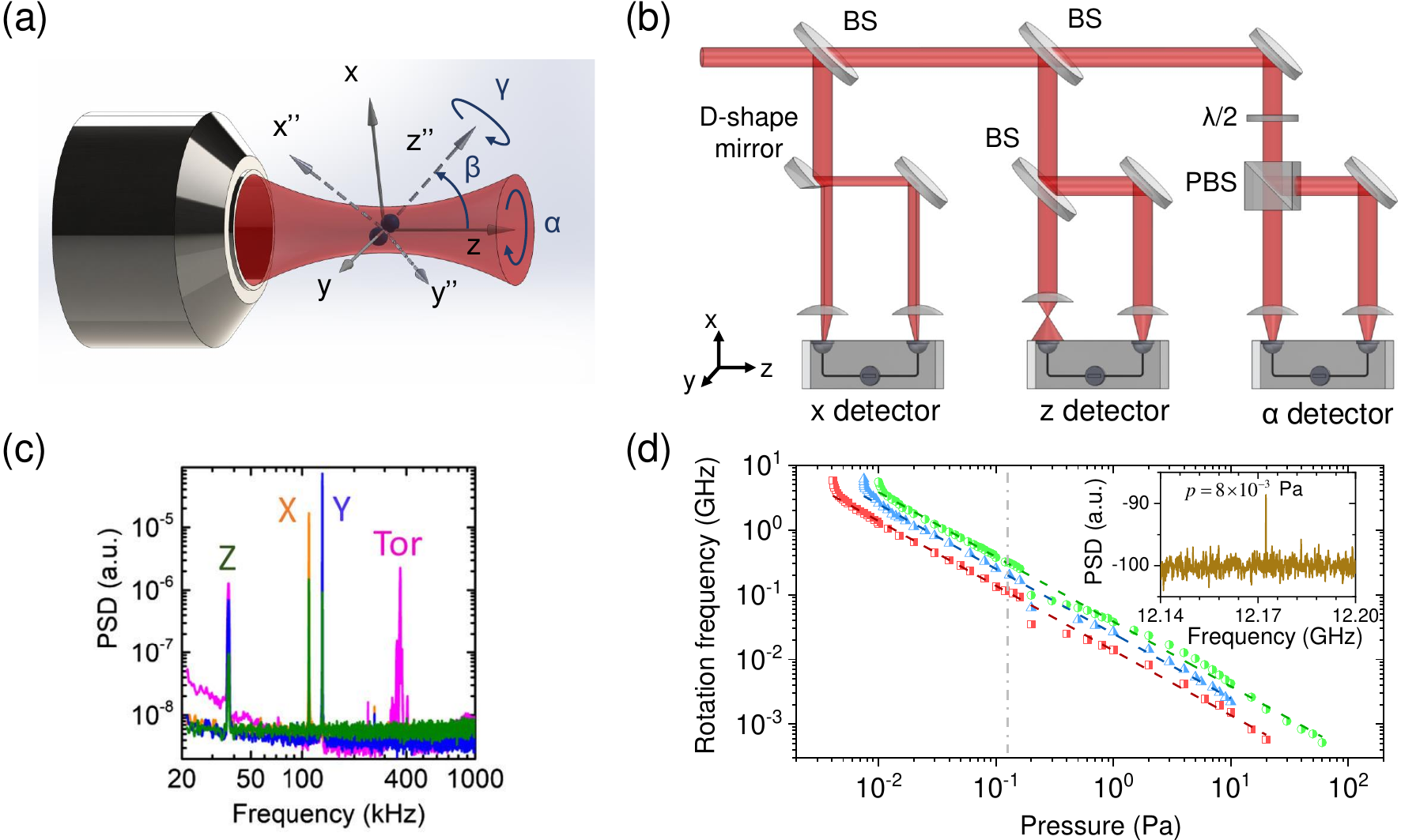}
	\caption{Trapping and detection of an optically levitated particle. (a) Schematic diagram of optical levitation and the multiple degrees of freedom, including  center-of-mass (CoM) motions along $x$, $y$, and $z$ axes, torsional vibration (libration) along $\alpha$ and $\beta$ directions, and free rotation along the $\gamma$ direction. (b) Schematic of detecting the CoM and rotational motions of a levitated particle. For x-motion detection, the collected laser is spatially separated by a D-shaped mirror. A balanced photodetector is used to measure the intensity difference between the two resulting beams, providing the x motion signal. The motion signals in the y and $\beta$ directions can be measured using the same method. For the measurement of the CoM motion in z direction, the beam is separated by a beam splitter (BS). The transmitted laser with strong power is partially collected in a photodiode, while the fully detected reflective laser (weak power) is used for comparison. The rotational motion in the $\alpha$ direction, which is polarization sensitive, can be measured using a half-wave plate ($\lambda/2$) and a polarized beam splitter (PBS) to separate the laser. (c) Power spectral density (PSDs) of the CoM motion in $x$, $y$ and $z$ directions and the torsional motion of a levitated particle. Reprinted with permission from Ref. \cite{Ahn2018Optically}, Copyright (2018) by the American Physical Society. (d) Rotation frequency in $\alpha$ direction of a levitated particle as a function of pressure. The inset is the PSD of the rotational motion at the pressure of $8 \times 10^{-3}$ Pa (equals to $7.5\times 10^{-6}$ Torr) \cite{Jin2021GHz}.}
	\label{fig:motion}
\end{figure*}

\subsection{Optical force}

Fig.~\ref{fig:motion}(a) shows a schematic diagram of the motion of an optically levitated particle. The optical force on a particle is influenced by optical scattering, a phenomenon dependents on both the particle's dimensions and the laser's wavelength. This force comprises two constituent components: the gradient force and the scattering force. The scattering force is aligned with the laser's propagation direction, while the gradient force is directed towards the focal point, collectively creating a three-dimensional trapping potential. 

In the Rayleigh regime, where the particle's radius $R$ is much smaller than the laser wavelength ($R \ll \lambda$), the particle is treated as a point dipole with Rayleigh approximation. The gradient force generated by the focused laser on the particle can be expressed as \cite{Harada1996Radiation}
\begin{equation}
	{F_\text{grad}}(x,y,z) = \frac{{2\pi {n_m}{R^3}}}{c}\left( {\frac{{{n^2} - 1}}{{{n^2} + 2}}} \right)\nabla I(x,y,z)
	\label{eq:gradient},
\end{equation}
and the scattering force is
\begin{equation}
	{F_\text{scat}}(x,y,z) = \frac{{128{\pi ^5}{n_m}{R^6}}}{{3c{\lambda ^4}}}{\left( {\frac{{{n^2} - 1}}{{{n^2} + 2}}} \right)^2}I(x,y,z)
	\label{eq:scattering},
\end{equation}
where $n_m$ is the refractive index of the surrounding environment, $n = n_0 / n_m$ is the relative refractive index of the particle, $c$ is the speed of light in vacuum. The intensity distribution of a Gaussian laser is given by $I(x,y) = \left[ {2P/\left( {\pi \omega _z^2} \right)} \right]\exp \left[ { - 2({x^2} + {y^2})/\omega _z^2} \right]$, where $P$ is the laser power, $\omega_z$ is the beam size along the laser propagation direction. % The beam waist is dependent on the numerical aperture (NA) of the focused laser, which is defined as ${\omega _0} = \lambda / \left[{\pi n_m \tan \left( {\arcsin \mathrm{NA}} \right)} \right]$. 
Generally, the gradient force is significantly larger than the scattering force and gravity of the particle, rendering the latter two forces negligible in the Rayleigh regime.

When the particle size is comparable to the laser wavelength ($R \sim \lambda$), the optical force on the particle can be determined using the Lorentz-Mie theory, which involves analyzing the solutions of electromagnetic fields surrounding the levitated particle. The optical force exhibits a dependence on the particle's shape and size \cite{Asano1979Light,Ren1997Scattering,Tzarouchis2018Light}, distinguishing it from the behavior observed in the Rayleigh regime. Additionally, when the particle size is significantly larger than the wavelength ($R \gg \lambda$), geometric optics becomes applicable for describing the optical force.

\subsection{Optical levitation schemes}

For nano-scale dielectric particles, such as silica nanoparticles, a single-beam trap generated by a tightly focused laser proves effective in stably levitating particles with a high trapping frequency, typically ranging from tens of kilohertz to a few megahertz \cite{Jin2018Optically,Jin2019Polarization,Shen2021Onchip,Magrini2021Real,Tebbenjohanns2021Quantum}. However, when particles are approximately 1 $\mu$m in size or have a high refractive index, such as silicon particles, the scattering force approaches or even exceeds the gradient force. In these cases, the particles are suitable for levitating with a dual-beam trap, which can be formed by two counter-propagating focused lasers \cite{Li2010Measurement,Li2013Fundamental,Yu2022Hermitian}. The scattering forces generated by the two counter-propagating beams are canceled due to the equal amplitudes and opposite directions. Consequently, laser power can be adjusted to lower levels, where the gradient force surpasses the gravity. Alternatively, two beams with the same optical frequency result in a standing wave, which can also be used for particle levitation \cite{Ranjit2016Zeptonewton,Kuhn2017Optically}. 
Despite the high trapping frequency along the laser propagation direction, the trapping frequency in the radial direction remains comparatively low, typically in the range of a few kilohertz \cite{Melo2023Vacuum}.

For large particles with masses on the order of nanograms, the gradient force may be inadequate to counteract gravity. An optical-gravitational trap, wherein the laser propagates in the direction opposing gravity, provides an alternative method for particle levitation. The scattering force acting on the particles is utilized to balance the gravitational force. Due to the substantial mass of the trapped particles, the resulting trapping frequency is low, typically in the range of tens of hertz \cite{Monteiro2020Force,Li2023Flexible}.

\subsection{Motion of levitated particles}

\begin{table*}
	\caption{Optically levitated particles and corresponding parameters. $m$: mass of particle; $\lambda$: wavelength of trapping laser; $P$: power of trapping laser; $f$: eigenfrequency of the motion; $p$: the operating pressure; $T_\text{eff}$: the cooling temperature.}
	\label{tbl:particles}
	\resizebox{\textwidth}{!}{%
		\renewcommand{\arraystretch}{1.2}
		\begin{tabular}{m{2.5cm}m{3.6cm}m{2cm}m{1.5cm}m{1.5cm}m{1.5cm}m{2cm}m{2cm}m{2cm}}
			\hline
			Particle & Shape & $m$ (kg) & $\lambda$ (nm) & $P$ (mW) & $f$ (kHz) & $p$ (Torr) & $T_\text{eff}$ (K) & Occupation \\
			\hline
			silica & nanosphere \cite{Delic2020Cooling} & $3 \times 10^{-18}$ & 1064 & 400 & 305 (x) & $8 \times 10^{-7}$ & $1.2 \times 10^{-5}$ & 0.43\textsuperscript{\emph{a}} \\
			& nanosphere \cite{Magrini2021Real} & $3 \times 10^{-18}$ & 1064 & 300 & 104 (z) & $7 \times 10^{-9}$ & - & 0.56\textsuperscript{\emph{b}} \\
			& nanosphere \cite{Tebbenjohanns2021Quantum} & $1 \times 10^{-18}$ & 1550 & 1200 & 77 (z) & $2 \times 10^{-9}$ & - & 0.65\textsuperscript{\emph{b}} \\
			& nanodumbbell \cite{Ahn2020Ultrasensitive} & $1 \times 10^{-17}$ & 1550 & 500 & 37 (z) & $1 \times 10^{-5}$ & - & - \\
			& nanodumbbell \cite{Bang2020Five} & $4 \times 10^{-18}$ & 1064 & 400 & 350 ($\alpha$) & $2 \times 10^{-3}$ & 9.2 & -\textsuperscript{\emph{c}} \\
			& nanoparticle \cite{Gao2024Feedback} & $2 \times 10^{-17}$ & 1550 & 900 & 334 ($\alpha$) & $3 \times 10^{-8}$ & $1.3 \times 10^{-3}$ & 84\textsuperscript{\emph{d}} \\
			& nanoellipsoidal \cite{Pontin2023Simultaneous} & $6 \times 10^{-18}$ & 1064 & 235 & 360 ($\beta$) & $4 \times 10^{-7}$ & $4.4 \times 10^{-3}$ & -\textsuperscript{\emph{a}} \\
			& nanoparticle \cite{Gieseler2012Subkelvin} & $3 \times 10^{-18}$ & 1064 & 100 & 37 (z) & $2 \times 10^{-6}$ & $5.0 \times 10^{-2}$ & -\textsuperscript{\emph{d}} \\
			& microsphere \cite{Li2011Millikelvin} & $3 \times 10^{-14}$ & 1064 & 220 & 9 (y) & $4 \times 10^{-5}$ & $1.5 \times 10^{-3}$ & 3400\textsuperscript{\emph{c}} \\
			& microsphere \cite{Li2023Flexible} & $4 \times 10^{-11}$ & 1064 & 500 & 0.1 (z) & $1 \times 10^{-7}$ & - & - \\
			silicon & nanorod \cite{Hu2023Structured} & $7 \times 10^{-17}$ & 1550 & 1200 & 150 (z) & 3 & - & - \\
			gold & nanosphere \cite{Jauffred2015Optical} & $5 \times 10^{-18}$ & 1064 & 340 & - & 760 & - & - \\
			vaterite & microsphere \cite{Arita2013Laser} & $1 \times 10^{-13}$ & 1070 & 25 & 0.42 (z) & $8 \times 10^{-3}$ & - & - \\
			polystyrene & microsphere \cite{Brzobohaty2023Synchronization} & $3 \times 10^{-16}$ & 1064 & 216 & - & 13 & - & - \\
			NaYF$_4$:Yb/Er & nanoparticle \cite{Zhang2023Determining} & $8 \times 10^{-18}$ & 1064 & 150 & 50 (z) & 2 & - & - \\
			& hexagonal prisms \cite{Winstone2022Optical} & $6 \times 10^{-15}$ & 1550 & 475 & 20 (z) & 2 & - & - \\
			YLF:Yb & nanoparticle \cite{Rahman2017Laser} & $2 \times 10^{-17}$ & 1031 & 200 & 52 (z) & 30 & - & - \\
			diamond & nanoparticle \cite{Hoang2016Electron} & $2 \times 10^{-18}$ & 1550 & 500 & 100 (z) & 9 & - & - \\
			\hline
		\end{tabular}%
	}
	\textsuperscript{\emph{a}} Cavity cooling;
	\textsuperscript{\emph{b}} Electric force feedback cooling;
	\textsuperscript{\emph{c}} Laser force feedback cooling;
	\textsuperscript{\emph{d}} Parametric feedback cooling.
\end{table*}

Considering the six degrees of freedom of a rigid object, we take a non-spherical particle, specifically a nanodumbbell (Fig.~\ref{fig:motion}(a)), as a representative example. The trapping laser, linearly polarized along the x direction, propagates in the z direction. The long axis of the levitated nanodumbbell tends to align with the laser polarization direction due to the interaction between the dipole moment and the electric field. The levitated particle undergoes Brownian motion in the potential well caused by collisions of gas molecules, involving three center-of-mass (CoM) motions in the x, y and z directions, two torsional motions denoted as $\alpha$ (in the $xy$-plane) and $\beta$ (in the $xz$-plane), as well as free rotation $\gamma$ around the long axis of the particle. 

The CoM motion equation of a levitated particle with a mass of $m$ in a one-dimensional harmonic trap can be expressed as \cite{Li2011Millikelvin,Li2013Fundamental}:
\begin{equation}
	\ddot x\left( t \right) + {\gamma _\text{CoM}}\,\dot x\left( t \right) + \Omega _0^2\,x\left( t \right) = {F_\text{th}}\left( t \right)/m
	\label{eq:motion},
\end{equation}
where $\Omega_0$ is the angular frequency of the CoM motion, $\gamma_\text{CoM}$ is the damping rate. The stochastic force caused by thermal noise is given by ${F_\text{th}}\left( t \right) = \sqrt {2{k_B}Tm\gamma_\text{CoM}} \delta \left( t \right)$, where $\left\langle {\delta \left( t \right)} \right\rangle  = 0$ and $\left\langle {\delta \left( t \right)\delta \left( {t'} \right)} \right\rangle  = \delta \left( {t - t'} \right)$. For a levitated sphere, the damping rate due to the collisions with gas molecules can be written as \cite{Li2011Millikelvin,Li2013Fundamental}
\begin{equation}
	\gamma_\text{CoM}  = \frac{{6\pi \eta R}}{m}\frac{{0.619}}{{0.619 + \text{Kn}}}\left( {1 + {c_k}} \right)
	\label{eq:damping},
\end{equation}
where the $\eta$ is the dynamic viscosity of the gas, Kn$= l/R$ is the Knudsen number, $l$ is the mean free path of the gas molecules, ${c_k} = 0.31\text{Kn}/\left( {0.785 + 1.152\text{Kn} + \text{Kn}^2} \right)$. Because of the large mean free path in high vacuum, $\text{Kn} \gg 1$, the damping rate retains the first-order terms given as ${\gamma_\text{CoM}} = 3.94\pi \eta {d^2}p/\left( {{k_B}T\rho R} \right)$, where $d$ is the mean diameter of the gas molecules. Based on the approximate expression, the damping rate is proportional to the pressure but inversely proportional to the particle's radius. The power spectral density (PSD) of the CoM motion is calculated by Eq:~\ref{eq:motion} with Fourier transformation \cite{Li2011Millikelvin},
\begin{equation}
	S\left( \omega  \right) = \frac{{2{k_B}T}}{m}\frac{{{\gamma _\text{CoM}}}}{{{{\left( {\Omega _0^2 - {\omega ^2}} \right)}^2} + {\omega ^2}\gamma _\text{CoM}^2}}
	\label{eq:psd},
\end{equation}
The rich information of CoM motion, including the trapping frequency and the damping rate, can be directly obtained from the PSD measurements.

The scattered light from a levitated particle, containing information about its motion, interferes with the trapping laser and can be measured by photodetectors, as shown in Fig.~\ref{fig:motion}(b). 
With the help of optical interference, the trapping laser amplifies the scattering light signal and improves the detection sensitivity of CoM motion extends to the order of $\mathrm{f m / \sqrt{Hz}}$.

Fig.~\ref{fig:motion}(c) shows the PSDs of the CoM and torsional motions of a 170-nm-diameter levitated silica nanodumbbell at a pressure of $5 \times 10^{-4}$ Torr \cite{Ahn2018Optically}. The left three peaks are the motion signals in the $z$ (38 kHz), $x$ (105 kHz), and $y$ (120 kHz) directions, respectively. The signal peak at 400 kHz is $\alpha$ torsional signal. The broad linewidth is induced by the coupling with the free rotation of the levitated particle around its long axis \cite{Bang2020Five}.

When the polarization of the trapping laser is switched to circular, it generates a torque on the levitated particle, inducing rotational motion around the $\alpha$ direction in vacuum. The rotation frequency of the levitated particle exhibits an inverse relationship with the pressure, as depicted in Fig.~\ref{fig:motion}(d). The inset is the PSD of the rotational motion at a pressure of $6 \times 10^{-5}$ Torr, indicating a rotation frequency about 6 GHz, which represents the fastest mechanical rotor demonstrated to date \cite{Jin2021GHz}.

The vacuum isolates the levitated particle from the physical contact with the external environment. 
%In low vacuum, a levitated particle remains stable due to air  damping. 
However, in high vacuum condition, the oscillation amplitude becomes larger and reaches the nonlinear trapping area without the air damping. Li \textit{et al.} first demonstrated experimental feedback cooling of the CoM motion of a levitated particle in 2011 \cite{Li2011Millikelvin}. This cooling method relies on force feedback utilizing the velocity signal of the levitated particle. Additionally, other methods, such as parametric feedback cooling and cavity cooling, have been reported to cool the motion of levitated particles. Remarkably, researchers have successfully cooled the CoM motion of levitated particles to the quantum ground-state \cite{Delic2020Cooling,Magrini2021Real,Tebbenjohanns2021Quantum}. Several typical cooling experiments are summarized in Table~\ref{tbl:particles}.

\subsection{Particles for optical levitation\label{sec:different_particles}}

Previous experiments have successfully demonstrated the stable optical trapping of various particle types. Based on the characteristics of levitated particles, the system reveals various applications, including fundamental physics studies, material science, and biological research. Table~\ref{tbl:particles} presents a  list of particles that have been optically levitated in air or vacuum conditions, along with their corresponding parameters.

Among numerous kinds of particles, silica particles are preferred in the experiments due to their small optical absorption coefficients and stability in vacuum and high temperature. Silica nanoparticles have been optically levitated in ultra-high vacuum and precisely controlled in six-dimensional motions \cite{Delic2020Cooling,Magrini2021Real,Tebbenjohanns2021Quantum,Bang2020Five,Gao2024Feedback,Pontin2023Simultaneous,Gieseler2012Subkelvin}. The systems exhibit high torque sensitivity with non-spherical particles, such as nanodumbbells \cite{Ahn2020Ultrasensitive,Ju2023Near}. Additionally, silica microparticles with masses on the nanogram scale have been successfully levitated using either dual-beam traps or optical-gravitational traps in vacuum, offering potential applications in inertial sensing and fundamental research \cite{Li2011Millikelvin,Li2023Flexible,Monteiro2020Force,Geraci2010Short,Chen2022constraining}.

Compared with the spherical or clustered shape of silica particles, fabricated silicon nanorods offer shape controllability, facilitating the study of the dynamics of levitated particles. The higher refractive index of silicon particles leads to significantly larger optical forces and torques under identical conditions \cite{Kuhn2017Optically,Hu2023Structured}. Moreover, birefringent particles, such as vaterite, experience both torque and translational force in an optical trap \cite{Arita2013Laser,Arita2020Coherent,Arita2023Cooling,zeng2024optically,Brzobohaty2023Synchronization}. NaYF:Yb/Er and YLF:Yb particles are specifically utilized to explore laser refrigeration \cite{Rahman2017Laser,LuntzMartin2021Laser,Winstone2022Optical,Zhang2023Determining}. In addition to dielectric particles, metal particles can also be employed in levitation systems to investigate different phenomena. Gold nanoparticles can be used to study surface plasmon resonance away from a surface to eliminate the effect of particle-surface interaction \cite{Demergis2012Ultrastrong,Jauffred2015Optical}. Notably, levitated diamonds with embedded NV centers, in contrast to classical particles, serve as highly sensitive quantum sensors, offering valuable insights into precise measurements \cite{Bose2017Spin,Marletto2017Gravitationally,Ledbetter2012Gyroscopes,Zhang2023Highly,Ma2017Proposal,Rusconi2022Spin} and fundamental researches \cite{Yin2013Large,Scala2013Matter,Delord2020Spin,Chudo2014Observation,Wood2017Magnetic,Maclaurin2012Measurable,Chen2019Nonadiabatic}. 

Apart from silica particles, the stability of other materials is compromised in high vacuum due to the heating induced by the trapping laser, which can be solved by using ion traps or magnetic traps. The multifaceted applications of optical levitation highlight its adaptability, wherein the selection of particles is finely tuned to the particular objectives of each experiment and the distinctive properties inherent to the particles themselves.

\section{Emerging applications}

Levitated optomechanics, which involves the detection and control of mechanical systems through optical forces, has sparked interest for its prospective real-world applications. Thanks to their isolation from the thermal environment, levitated particles in vacuum exhibit ultra-high sensitivity, enabling precise measurements of weak fields. Herein, we present a few potential real-world applications that are relevant to our daily lives.

\subsection{Ultrasensitive force and torque detection}

\begin{figure*}
	\includegraphics[width=0.9\textwidth]{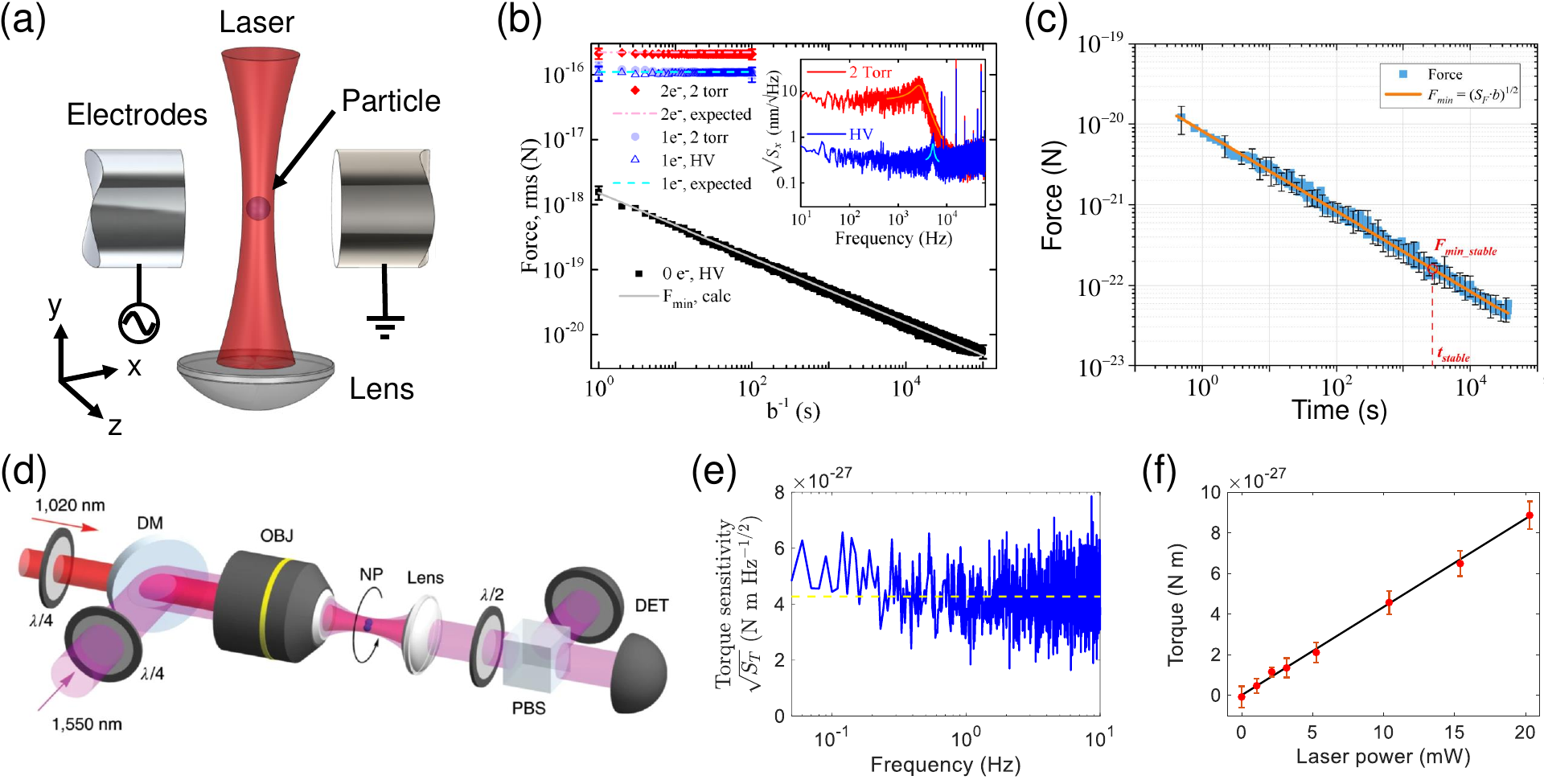}
	\caption{Ultrasensitive force and torque detection with optically levitated particles. (a) Schematic diagram of the experimental setup for force measurement. A particle is levitated in an optical trap, exhibiting high sensitivity to external force. A sinusoidal electric field generated by a pair of electrodes is used to calibrate the force acting on the particle. (b) Measured force on charged and uncharged nanoparticles in an electric field as a function of measurement time. Reprinted with permission from Ref. \cite{Ranjit2016Zeptonewton}, Copyright (2016) by the American Physical Society. The minimum detectable force is $5.8 \pm 1.3$ zN for averaging times exceeding $10^5$ s, which corresponds to a force sensitivity of $1.6 \pm 0.37$ $\mathrm{aN/\sqrt{Hz}}$. (c) Minimum resolvable force as a function of measurement time by Liang \textit{et al.} \cite{Liang2023Yoctonewton}. The minimum detectable force is $40.80 \pm 8.55$ yN and the corresponding force sensitivity is $6.33 \pm 1.62$ $\mathrm{zN/\sqrt{Hz}}$. (d) Experimental setup for optically levitated particles and the external torque measurement designed by Ahn \textit{et al.} \cite{Ahn2020Ultrasensitive}. A silica nanoparticle is levitated in vacuum using a 1550 nm laser focused by an objective lens. An external torque is applied on the nanoparticle using a 1020 nm laser. (e) Demonstrating ultra-high torque sensitivity of $4.2 \times 10^{-27}$ $\mathrm{N m / \sqrt{Hz}}$ for the levitated particle at a pressure of $1.3 \times 10^{-5}$ Torr. (f) External torque measurements generated by the 1020 nm laser at varying power levels \cite{Ahn2020Ultrasensitive}.}
	\label{fig:forcetorque}
\end{figure*}

Levitated particles in high vacuum experience minimal thermal noise and friction due to reduced collisions with gas molecules, resulting in a high quality factor ($Q$) and holding promise for precise force measurements \cite{Ranjit2015Attonewton,Ranjit2016Zeptonewton,Hempston2017Force,Hebestreit2018Sensing,Liang2023Yoctonewton}. The force sensitivity of a levitated harmonic oscillator limited by the thermal noise is given by
\begin{equation}
	S_F^{1/2} = \sqrt {4{k_B}Tm{\Omega _0}/Q}
	\label{eq:forcesensitivity},
\end{equation}
where $T$ is the motion temperature, $\Omega _0/Q = \gamma_\text{CoM}$ is the damping rate of the CoM motion. Therefore, the minimum resolvable force during a measurement time of $t$ is ${F_{\min }} = \sqrt {4{k_B}Tm{\gamma _\text{CoM}}/t}$. 

One of the challenges in precise force measurement is the calibration. Generally, the force can be calibrated with a known electric field, as shown in Fig.~\ref{fig:forcetorque}(a). An optically levitated particle is positioned in an alternating current (AC) electric field, produced by a pair of electrodes. The displacement of the levitated particle under a sinusoidal electric field is denoted as $x\left( t \right) = \left( {qE/m{\omega _\text{ele}}{Z_m}} \right)\sin \left( {{\omega _\text{ele}}t + \varphi } \right)$. Here, $q$ is the charge on the particle, $E$ and $\omega _\text{ele}$ are the amplitude and frequency of the electric field, ${Z_m}\left( {{\omega _\text{ele}}} \right) = {\left[ {{{\left( {\Omega _0^2 - \omega _\text{ele}^2} \right)}^2} + \gamma _\text{CoM}^2} \right]^{1/2}}$ is the impedance of the oscillation at the frequency of $\omega_\text{ele}$, and $\varphi$ is the phase of the oscillation relative to the electric field. The charge on the levitated particle can be conveniently measured and controlled through electric discharge \cite{Frimmer2017Controlling} or ultraviolet (UV) light \cite{Ashkin1980Applications}. Consequently, the relationship between the force and the motion signal can be accurately determined. This method also enables the precise measurement of the mass of the levitated particle.

Ranjit \textit{et al.} have demonstrated a force sensitivity of $(1.6 \pm 0.37) \times 10^{-18}$ $\mathrm{N/\sqrt{Hz}}$ with optically levitated silica nanospheres \cite{Ranjit2016Zeptonewton}. The minimum measurable force is a function of measurement time, as illustrated in Fig.~\ref{fig:forcetorque}(b), with a value of $(5.8 \pm 1.3 ) \times 10^{-21}$ N achievable for measurement time exceeding $10^5$ s. Recently, the minimum resolvable force of $(40.80 \pm 8.55) \times 10^{-24}$ N was achieved by Liang \textit{et al.} (Fig.~\ref{fig:forcetorque}(c)) \cite{Liang2023Yoctonewton}, with a force sensitivity of $(6.33 \pm 1.62)  \times 10^{-21}$ $\mathrm{N/\sqrt{Hz}}$.

In addition, levitated particles can be driven to rotate w\cite{Bang2020Five,Pontin2023Simultaneous} and indicate ultra-high torque sensitivity \cite{Ahn2020Ultrasensitive,Ju2023Near}. 
% Duo to the high torque sensitivity in vacuum \cite{Ahn2020Ultrasensitive,Ju2023Near}, levitated rotors are suitable for numerous precise measurements. 
In the case of an optically levitated rotor, driven by a circularly polarized trapping laser, the rotational motion is determined by both the driving torque ($M_\text{opt}$) exerted by the trapping laser and the damping torque ($M_\text{gas}$) arising from the surrounding gaseous environment. The angular rotation frequency $\omega_r$ satisfies $Id{\omega _r}/dt = {M_\text{opt}} + {M_\text{gas}}$, where $I$ is the moment of inertia, which is $I = 0.4 m {R^2}$ for a spherical particle with a radius of $R$. The laser driving torque originates from three components, including absorption, birefringence and shape asymmetry of the levitated particle \cite{Ahn2018Optically}. All three components are proportional to the laser intensity. 
% A levitated particle starts to rotate and accelerate due to the driving torque, 
During the rotational acceleration of the levitated rotor, the damping torque (${M_\text{gas}} = - I{\omega_r}{\gamma_\text{rot}}$) increases simultaneously, while the driving torque remaining constant. Here, $\gamma_\text{rot}$ is the damping rate of the rotational motion. Finally, when the sum of driving and damping torques is zero, and the rotation frequency ends at ${\omega_r} = {M_\text{opt}}/I{\gamma_\text{rot}}$.

% In measurements, an external torque acts on the levitated particle, leading to a modification in its rotational motion. 
Fig.~\ref{fig:forcetorque}(d) shows a schematic of torque measurement with optically levitated particles \cite{Ahn2020Ultrasensitive}. An external torque generated by a 1020 nm laser is applied on the levitated particle, which can be measured by monitoring the change of the rotation frequency. Similarly, the torque sensitivity is limited by the thermal noise induced by the residual gas molecules given by
\begin{equation}
	S_M^{1/2} = \sqrt {4{k_B}TI{\gamma_\text{rot}}}
	\label{eq:torquesensitivity}.
\end{equation}

\begin{figure*}
	\includegraphics[width=1\textwidth]{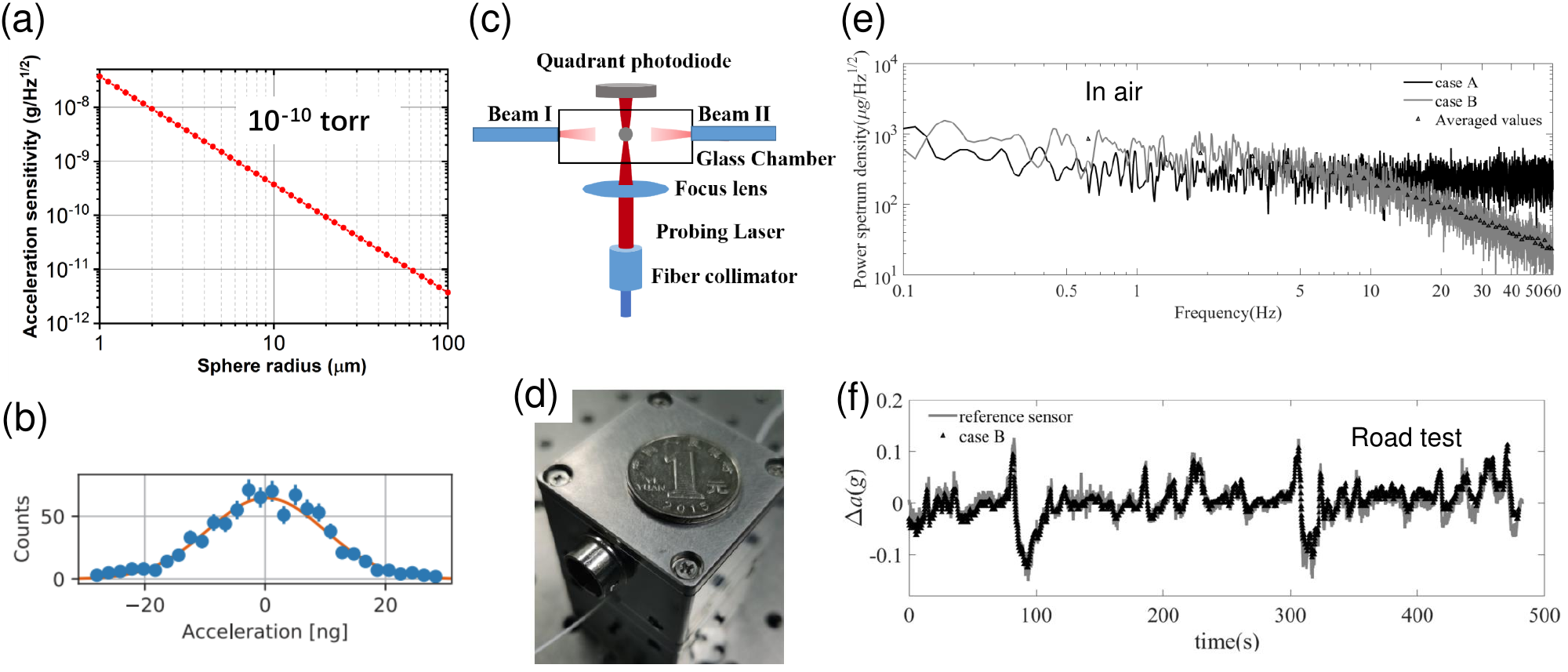}
	\caption{Sensitivity of levitated optomechanical accelerometers. (a) Theoretical calculation of acceleration sensitivity as a function of levitated particle radius at a pressure of $10^{-10}$ Torr. (b) Acceleration measurements during 52 s integration segments over a total of 12 h using optically levitated massive particles in the absence of external force. Reprinted with permission from ref \cite{Monteiro2020Force}, Copyright (2020) by the American Physical Sociey. Monteiro \textit{et al.} demonstrated an observable acceleration of $170 \pm 340 \pm 70$ pg. (c) Schematic and (d) optical image of a small accelerometer designed by Han \textit{et al.} \cite{Han2023Feedback}. A silica microparticle is levitated by a dual-beam trap using two optical fibers with counter-propagating lasers.  (e) Acceleration sensitivity spectrum of the accelerometer without (black curve) and with (gray curve) feedback cooling of CoM motion. The sensitivity of $232 \pm 120$ $\mathrm{\mu g/\sqrt{Hz}}$ in the frequency range of 1-10 Hz and fall down to $25.5 \pm 8.2$ $\mathrm{\mu g/\sqrt{Hz}}$ at the frequency of 55 Hz have been measured. (f) Performance of the accelerometer installed on a vehicle running on a road.}
	\label{fig:sensitivity}
\end{figure*}

The torque sensitivity spectrum at a pressure of $1.3 \times 10^{-5}$ Torr and room temperature is shown in Fig.~\ref{fig:forcetorque}(e). The highest sensitivity of $\left( 4.2 \pm 1.2 \right) \times 10^{-27}$ $\mathrm{N m / \sqrt{Hz}}$ is experimentally demonstrated  \cite{Ahn2020Ultrasensitive}, which greatly surpasses that of state-of-the-art nanofabricated torque sensors at millikelvin temperatures ($2.9 \times 10^{-24}$ $\mathrm{N m / \sqrt{Hz}}$) \cite{Kim2016Approaching}. The torque measurements for various 1020 nm laser power are presented in Fig.~\ref{fig:forcetorque}(f). The measured external torque is as low as $4.3 \times 10^{-28}$ Nm over a measurement duration of 100 s with a modulation power of 1.1 mW. Such high torque sensitivity enables the levitated particles to offer potential applications in detecting vacuum friction and Casimir torque \cite{Kardar1999The,Zhao2012Rotational,Xu2021Enhancement,Xu2017Detecting,Manjavacas2017Lateral,Ju2023Near}.

\subsection{Inertial sensing: accelerometers and gyroscopes} 

Acceleration measurements are essential across various industries and wildly used in automotive, health and fitness, aerospace and aviation, and robotics. 
% This highlights the versatility of acceleration measurement technology. 
A few technologies are applied to measure acceleration, consisting of piezoelectric accelerometers \cite{Yang2010A}, MEMS (Micro-Electro-Mechanical Systems) accelerometers \cite{Alessandro2019A}, optical accelerometers \cite{Lu2021Review}, and so on. LIGO (Laser Interferometer Gravitational-Wave Observatory) instruments, as extremely sensitive detectors for measuring gravitational waves, achieve a sensitivity at the level of $10^{-10}$ $\mathrm{g/\sqrt{Hz}}$ (where $g = 9.8\, \mathrm{m / s^2}$) \cite{Hines2023Compact}. However, the impressive sensitivities are realized under controlled laboratory settings and typically limited to specific frequency ranges. The substantial size and significant cost of LIGO make it impractical for real-world applications. 

In contrast, levitated optomechanics, utilizing micro- or nano-particles, has been demonstrated acceleration sensitivities at the order of $10^{-7}$ $\mathrm{g/\sqrt{Hz}}$ \cite{Monteiro2020Force}. Theoretically, Fig.~\ref{fig:sensitivity}(a) shows calculated acceleration sensitivity as a function of the particle size at a pressure of $10^{-10}$ Torr. The sensitivity can reach $3 \times 10^{-12}$ $\mathrm{g/\sqrt{Hz}}$ with a 100 $\mu$m particle. Compared to smaller particles, levitated massive particles demonstrate heightened sensitivity to acceleration but diminished sensitivity to force \cite{Gonzalez2021Levitodynamics}. Given their notable sensitivity and cost-effectiveness, levitation systems hold significant potential across various applications, including the detection and monitoring of seismic activities, precise measurement of acceleration during flight, and monitoring the structural integrity of buildings, bridges, and other infrastructure.

Fig.~\ref{fig:forcetorque}(a) shows the schematic diagram for acceleration measurements. A  massive particle ($\sim$ ng) is levitated in an optical-gravitational trap, formed by a vertically propagating 1064 nm laser focused by a low NA lens with a long working distance. Monteiro \textit{et al.} demonstrated an acceleration sensitivity of $95 \pm 41$ $\mathrm{ng/\sqrt{Hz}}$ at frequencies near 50 Hz using optically levitated silica microspheres with a diameter of 10 $\mu$m \cite{Monteiro2020Force}. The corresponding force sensitivity is $0.95 \pm 0.11$ $\mathrm{aN / \sqrt{Hz}}$. Fig. \ref{fig:sensitivity}(b) gives the distribution of the accelerations measured during 52 s integration segments over a total of 12 hours in the absence of external force. Based on the Gaussian fitting, a minimum observable acceleration is $170 \pm 340\,\text{[stat]} \pm 70\,\text{[syst]}$ pg. 

In addition to measuring AC acceleration and force, \cite{Ranjit2016Zeptonewton,Monteiro2020Force}, levitated optomechanics can also detect static forces using free-falling nanoparticles, such as gravity and static electric forces \cite{Hebestreit2018Sensing}. The force sensing scheme consists of three steps: (i) A particle is trapped in a harmonic potential with a high stiffness coefficient, sufficiently large to neglect the displacement of the levitated particle induced by the static force. (ii) The trapping potential is turned off for an interaction time. Under the influence of static force, the particle undergoes displacement, which depends on the amplitude of acceleration and time duration. (iii) Restarting the trapping potential produces an increased amplitude oscillation at the resonance frequency compared to the initial state. At high vacuum, the amplitude of displacement can be precisely measured to calculate the static force. Feedback cooling of the CoM motion can be applied to decrease the initial velocity of the levitated particle and improve measurement precision. Hebestreit \textit{et al.} demonstrated a sensitivity of 10 aN for measuring static gravitational and electric forces \cite{Hebestreit2018Sensing}.

The above-mentioned experiments of sensitivity measurements with levitation systems are carried out in lab condition. To facilitate their practical applications, levitation systems have been packed as a small-scale accelerometer. The schematic and optical image of an accelerometer with a size on the centimeter scale designed by Han \textit{et al.} are shown in Fig. \ref{fig:sensitivity}(c) and (d) \cite{Han2023Feedback}. A silica microsphere is levitated in a dual-beam optical trap in air. Fig. \ref{fig:sensitivity}(e) is the measured acceleration sensitivity spectrum of the accelerometer without (black curve) and with (gray curve) feedback cooling, demonstrating a sensitivity of $25.5 \pm 8.2$ $\mathrm{\mu g/\sqrt{Hz}}$ at the frequency of 55 Hz. Furthermore, the performance of the packed sensor is evaluated by testing it on a vehicle running on a road (Fig. \ref{fig:sensitivity}(f)), providing an example of real-world applications. 

As inertial sensors, levitated optomechanics can act as gyroscopes for measuring orientation, angular velocity, and torque. Traditional gyroscopes rotate freely in all directions, which is a crucial characteristic for maintaining orientation. Beyond rotor gyroscopes that operate based on the conservation of angular momentum, various alternative types exist, such as the ring laser gyroscope or fiber optic gyroscope \cite{Lee2003Review}, relying on the Sagnac effect, and MEMS gyroscopes \cite{Tadigadapa2009Piezoelectric}, using the Coriolis force. Gyroscopes have diverse applications, serving in navigation systems for aircraft, inertial measurement units in robotics and smartphones, stabilization systems in cameras and various other fields. MEMS gyroscopes are prevalent in portable electronic devices due to their compact size and low power consumption. However, the sensitivity of MEMS gyroscopes is relatively low. Instead, laser gyroscopes and fiber optic gyroscopes employ laser beams and fiber optics to measure orientation changes, offering high sensitivity and commonly used in aerospace and high-precision applications. Nonetheless, their substantial size and high cost restrict their application range. 

Levitated optomechanics is a promising alternative for gyroscopes, combining the advantages of high sensitivity, small size, and low cost. Optically levitated rotors have been reported ultra-high torque sensitivity \cite{Ahn2020Ultrasensitive,zeng2024optically}. Moreover, Zhang \textit{et al.} proposed a scheme based on an NV center in a levitated nanodiamond to measure the angular
velocity through matter-wave interferometry with an ultra-high sensitivity of $6.86 \times 10^{-7}$ $\mathrm{rad / s / \sqrt{Hz}}$ in an ion trap \cite{Zhang2023Highly}.

\subsection{Electric and magnetic field sensing} 

\begin{figure}
	\includegraphics[width=0.5\textwidth]{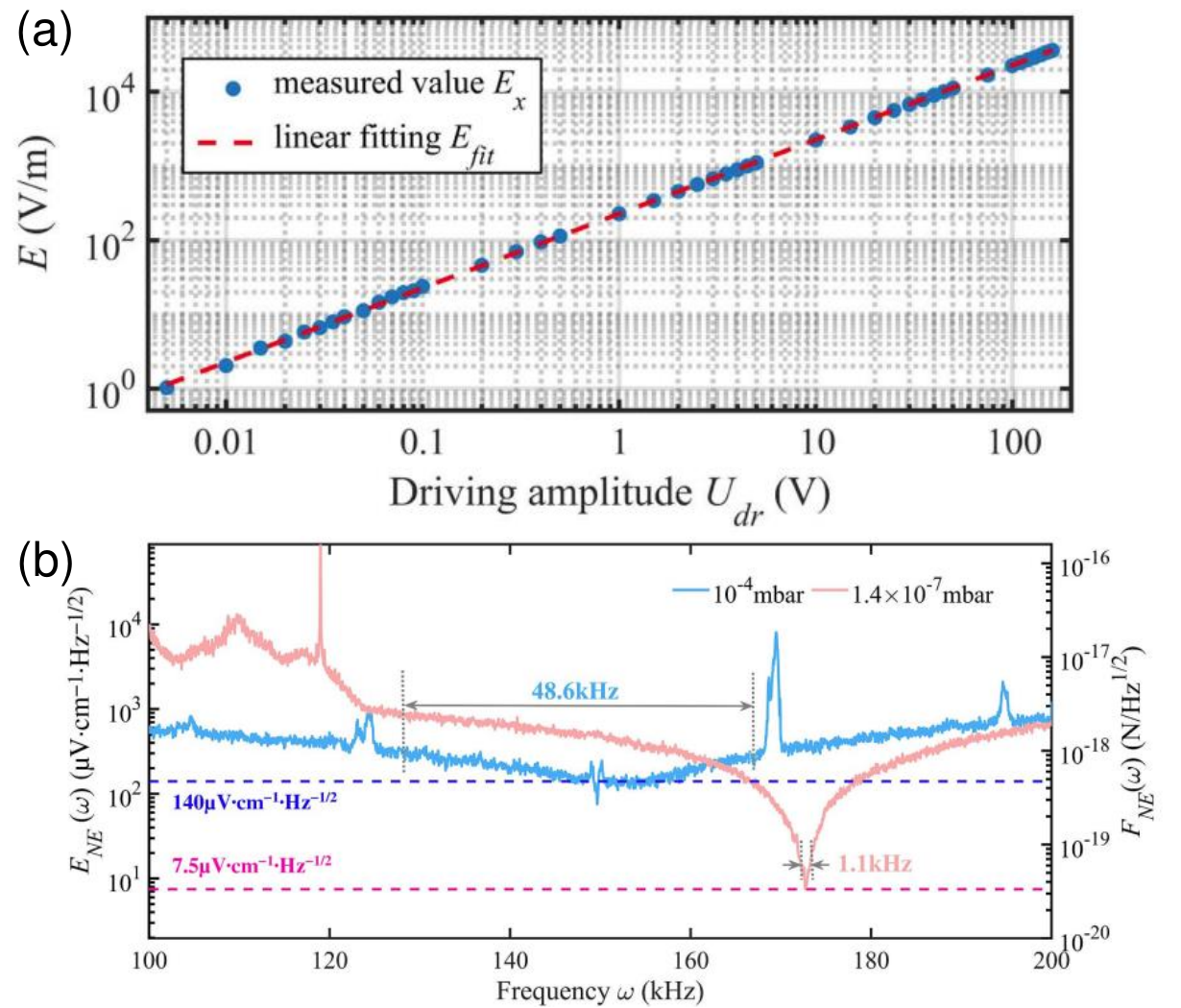}
	\caption{Electric field sensing using levitated optomechanics \cite{Zhu2023Nanoscale}. (a) Electric field measurements at various voltage on parallel electrodes with an optically levitated charged particle. The measured electric field varies from 1.03 V/m to 36.2 kV/m, which spans over 4 order of magnitudes. (b) Noise equivalent force and equivalent electric field. The sensitivity reaches to 7.5 $\mathrm{\mu V / cm / \sqrt{Hz}}$ at $1.4 \times 10^{-7}$ mbar (equal to $1.05\times 10^{-7}$ Torr).}
	\label{fig:electric}
\end{figure}

In the presence of an external field, such as an electric or magnetic field, the levitated particle will experience a force or torque if the particle is charged or has a magnetic moment.  As introduced in previous section, levitation systems exhibit an ultra-high force and torque sensitivity to external perturbations.  Consequently, levitated optomechanics can serve as weak field sensors for precision measurement. Compared with traditional sensing techniques, the non-contact characteristic of levitation systems removes the need for  electrodes with physical contacts, enabling measurements in a wide range of environments, 
%. Electric and magnetic field sensing are wildly used in environment monitoring and various engineering, 
including meteorology, industrial automation, biomedical engineering and telecommunications.

\begin{figure*}
	\includegraphics[width=0.9\textwidth]{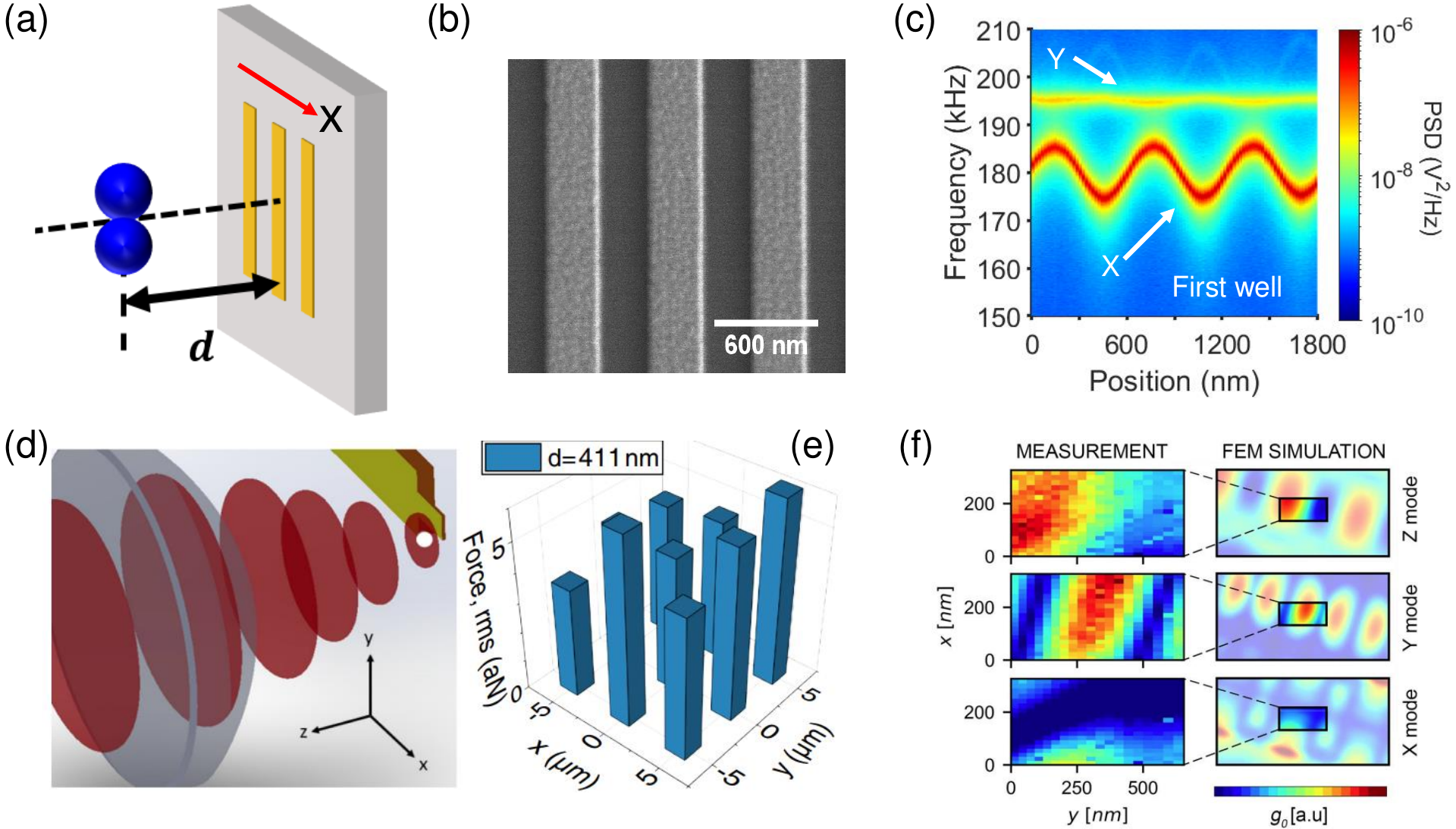}
	\caption{Probe microscopy of Levitated Optomechanics. (a) Schematic diagram of levitated scanning probe microscopy using a optically levitated neutral particle near a surface \cite{Ju2023Near}. The surface features a nanograting structure along the x direction. (b) SEM image of the gold nanograting structure with a width of 300 nm. The scale bar corresponds to a length of 600 nm. (c) PSDs of the levitated particle scanning along the x-direction when the particle-to-surface distance is approximately 370 nm. The trapping frequency of the levitated particle in the x direction exhibits periodic changes, while the trapping frequency in the y direction remains constant. (d) Schematic of scanning force sensing with a levitated nanosphere in an optical lattice \cite{Montoya2022Scanning}. The particle is trapped at micrometer distances from a conductor surface. (e) Force map measured at the resonance frequency of CoM motion at a distance of 0.411 $\mu$m with a measurement time of 20 s. The force sensitivity is independent on the distance. (f) Measured intensity map of the single-photon optomechanical coupling rates using a levitated probe near a nanocavity structure \cite{Magrini2018Near}. (d) and (e) are adapted with permission from ref \cite{Montoya2022Scanning} \copyright Optical Society of America.} 
	\label{fig:probe}
\end{figure*}

Using levitated dielectric nanoparticles with a certain net charge, Zhu \textit{et al.} demonstrated three-dimensional electric field measurement with a high sensitivity \cite{Zhu2023Nanoscale}. By scanning the nanoparticle position relative to the electric field generated by a pair of electrodes, the electric field distribution is obtained. The electric field strength with a range from 1.03 V/m to 36.2 kV/m is detected during a measurement time of 1 second by changing the voltage on the electrodes, as shown in Fig.~\ref{fig:electric}(a). The noise equivalent electric field reaches 7.5 $\mathrm{\mu V / cm / \sqrt{Hz}}$ at $1.4 \times 10^{-7}$ mbar (equal to $1.05\times 10^{-7}$ Torr, Fig.~\ref{fig:electric}(b)). Recently, Fu \textit{et al.} proposed a prototype that uses levitated nanoparticles to measure AC electric field and serve as low-frequency receiving antennas \cite{fu2024optically}.

Similarly, levitated ferromagnetic particles can be used to sense magnetic fields \cite{JacksonKimball2016Precessing,Jiang2020Superconducting,Ahrens2024Levitated}. A ferromagnetic particle undergoes rotation due to the induced torque arising from the interaction between its intrinsic magnetic moment and an external magnetic field. Jiang \textit{et al.} designed a superconducting levitation system to trap a neodymium magnetic disk attached to a high-reflectivity mirror, forming a Fabry-P\'{e}rot cavity in combination with another mirror. This design allows for precise detection of the dynamics of the levitated magnet under the interaction with external magnetic field, demonstrating a sensitivity in magnetic field measurements of 370 $\mathrm{pT / \sqrt{Hz}}$ \cite{Jiang2020Superconducting}. Such a system for magnetometry even exceeds the standard quantum limit ($\hbar$) of magnetic field measurement by quantum magnetometers, including superconducting quantum interference devices, solid-state spins, and optical atomic magnetometers. Recently, Ahrens \textit{et al.} reported a sensitivity of 20 $\mathrm{fT / \sqrt{Hz}}$, corresponding to 0.064 $\hbar$, with levitated ferromagnetic particles in a superconducting trap \cite{Ahrens2024Levitated}. These highly sensitive systems provide a powerful tool for precision measurements and testing fundamental physics.

\subsection{Levitated scanning probe microscopy}

In the realms of materials, biology and manufacturing, the detection of surface structures is crucial for analyzing sample properties, structures, and quality. Various techniques are used for the detection, including optical microscopy, SEM, and atomic force microscopy (AFM). The resolution of optical microscopy is restricted by the diffraction limit, which is determined by the wavelength of light and the NA of objective lens, resulting in a resolution capped at a few hundred nanometers with visible lights. In contrast, SEM and AFM own higher resolutions in the nanometer range but are associated with drawbacks such as larger equipment size and higher costs. Levitated optomechanics emerge as highly sensitive probes enabling the study of surface characteristics with exceptional resolution.

Ju \textit{et al.} \cite{Ju2023Near} reported a neutral silica particle optically levitated near a sapphire surface featuring a nanograting structure, as shown in Figs.~\ref{fig:probe}(a) and (b). The interference between the reflected light from the surface and the trapping beam forms a standing wave. The particle can be stably trapped in the anti-nodes of the standing wave, with the first well situated approximately 370 nm away from the surface. Through systematic scanning of the nanograting with the levitated particle, the trapping frequency of the CoM motion varies periodically, as shown in Fig.~\ref{fig:probe}(c). Such a scanning method proves to be an effective tool for detecting and characterizing the surface structure beyond the diffraction limit.

Additionally, the force sensitivity of levitated optomechanics is not affect by surface structure or the separation between surface and particles. Montoya \textit{et al.} demonstrated a  silica nanosphere levitated by a standing wave near a conductor surface (Fig.~\ref{fig:probe}(d)) \cite{Montoya2022Scanning}. The minimum observable force measured at the resonance frequency of CoM motion with a distance of 0.411 $\mu$m is shown in Fig.~\ref{fig:probe}(e). By changing the distance of the levitated particle from the surface, no significant difference of force sensitivity is observed \cite{Montoya2022Scanning,Ju2023Near}.

The three-dimensional intensity gradient of a nanophotonic cavity  can also be imaged by a levitated particle, which has been achieved by Magrini \textit{et al.} \cite{Magrini2018Near}. A nanoparticle is levitated at a distance of 310 nm from a nanophotonic cavity using a stranding wave. Due to the coupling to the evanescent field of the cavity mode, the motion signal of the particle is affected by the phase fluctuations of the cavity mode. The optomechanical coupling rates between the particle and nanocavity is shown in Fig.~\ref{fig:probe}(f). Compared to AFM, the imaging resolution of a levitated particle is only limited by the measurement of the particle motion, rather than the size of the probe, which may reach up to tens of nanometers.

\subsection{Localized vacuum pressure gauge}

\begin{figure*}
	\includegraphics[width=0.9\textwidth]{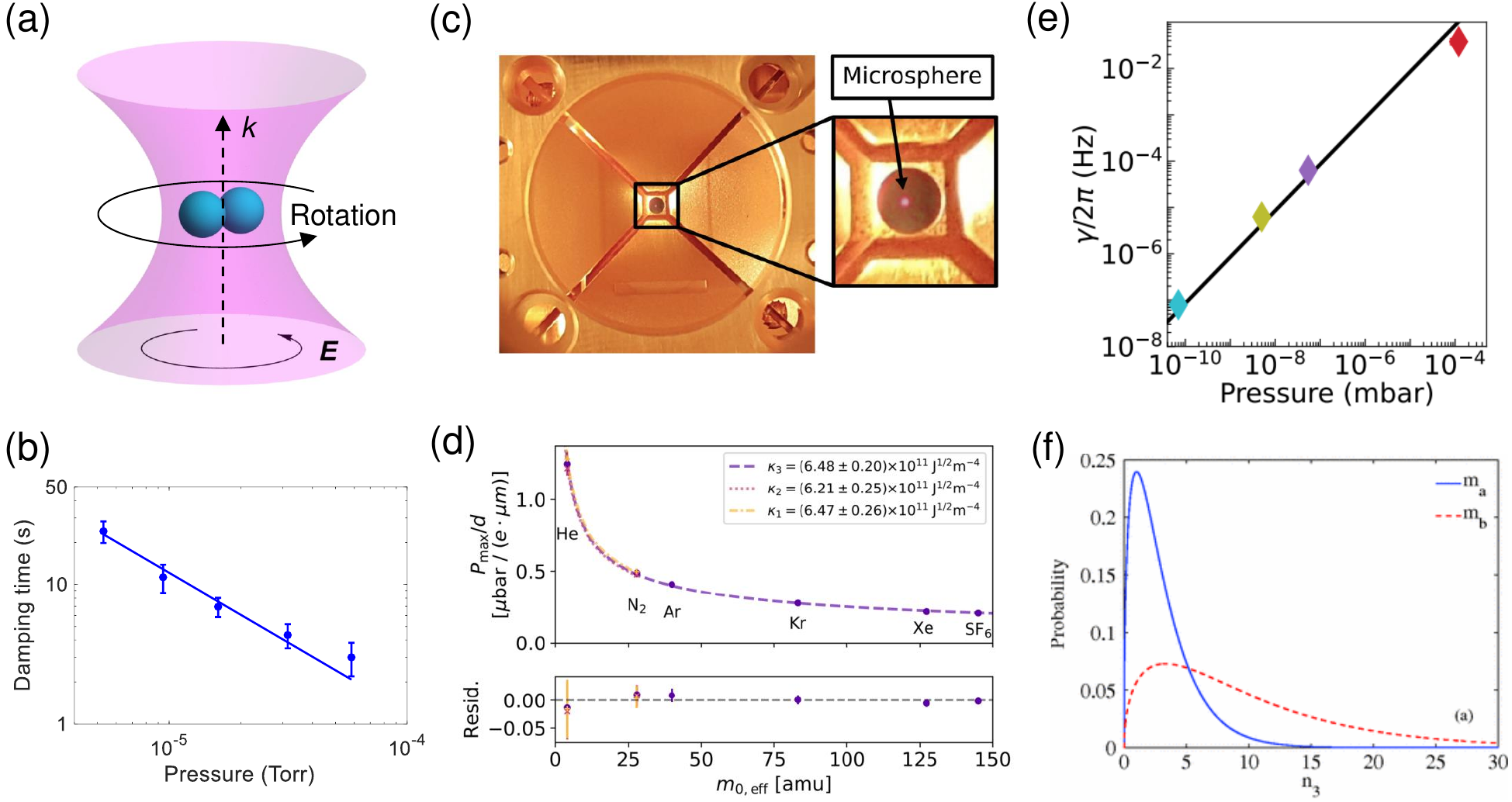}
	\caption{Vacuum gauge utilizing levitated optomechanics. (a) Optically levitated particle trapped and manipulated by a circularly polarized laser, inducing rotation in vacuum. Reprinted with permission from Ref. \cite{Ahn2018Optically}, Copyright (2018) by the American Physical Society. (b) Damping time of the rotational motion of a levitated particle as a function of pressure due to the collisions of gas molecules \cite{Ahn2020Ultrasensitive}. The damping time is inversely proportional to pressure. (c) Schematic of a optically levitated microsphere in the inner region of the shielding electrodes. The levitated particle rotates in an electric field generated by four electrodes \cite{Blakemore2020Absolute}. (d) Measured gas pressure using the levitated particle dependent on the effective mass of the gas molecules for different species spanning the range from 4 to 150 amu \cite{Blakemore2020Absolute}. (c) and (d) are reprinted with permission from Ref \cite{Blakemore2020Absolute}, Copyright (2020) American Vacuum Society. (e) Damping rate of the CoM motion of an optically levitated particle as a function of pressure. Reprinted with permission from Ref. \cite{Dania2023ultrahigh}, Copyright (2023) by the American Physical Society. (f) Theoretical calculation of the mean phonon increase of a ground-state cooling particle induced by the collision of a signal molecule. Reprinted with permission from ref \cite{Yin2011Three}, Copyright (2011) by the American Physical Society.} 
	\label{fig:gauge}
\end{figure*}

An ultra-high vacuum environment, approximately below  $10^{-8}$ Torr, is essential for various scientific research applications, including particle accelerators \cite{Wiedemann2015Particle}, gravitational wave detectors \cite{Harry2010Advanced}, thin film growth and preparation \cite{Chambers2000Epitaxial}, electron-beam lithography \cite{Vieu2000Electron}, atomic force microscopy \cite{Giessibl2003Advances}, and so on. Ionization gauges, including both hot and cold cathode types, represent the most sensitive methods in low-pressure measurement. These gauges collect the ionized gas molecules and measure the weak current, which is proportional to the gas pressure. 
% However, in extremely high vacuum conditions with ultra-low gas molecules density, the production of ions is limited and dependent on the nature of gases, which are typically mixed and not always well-known.
However, in extremely high vacuum conditions with ultra-low gas molecules density, the production of ions is limited and dependent on the types of gases, which are typically mixed and unknown.  Consequently, ionization gauges face challenges in calibration and large errors. In general, the pressure measurement range is limited to the magnitude of  $10^{-11}$ Torr. Plenty of specific researches require an even higher vacuum level, which cannot be precisely measured by conventional ion gauge.

Levitation system is a potential tool for measuring ultra-low pressure with high accuracy. The principle of such a vacuum gauge is based on the rotational motion of levitated particles. The damping torque caused by the collisions of gas molecules is given by ${M_\text{gas}} = - I{\omega_r}{\gamma_\text{rot}}$. Specifically, for a spherical particle, the damping rate of the rotational motion is ${\gamma_\text{rot}} = 10 \kappa p / \left( \pi \rho R v \right)$ \cite{Fremerey1982Spinning}, where $\kappa \approx 1$ is the accommodation factor of angular momentum transfer from gas molecules onto the particle, $\rho$ is the particle density, $p$ is the pressure, and $v = \sqrt {8{k_B}T/\pi {m_\text{gas}}}$ is the mean speed of gas molecules, $m_\text{gas}$ is the mass of a single gas molecule. Consequently, the damping rate of rotational motion is proportional to the gas pressure, implying that pressure can be obtained by measuring the damping rate.

The damping rate of the rotational motion of a levitated particle can be detected through various methods, which is based on distinct driving diagrams. When the particle is driven by the trapping laser (Fig.~\ref{fig:gauge}(a)), the rotation frequency is dependent on the ellipticity of the laser polarization \cite{Jin2021GHz}. After abrupt alternation of the laser polarization, the evolution of the rotation frequency can be expressed as a function of time ${\omega _r}\left( t \right) = {\omega _1} + \left( {{\omega _2} - {\omega _1}} \right)\left( {1 - {e^{ - \left( {t - {t_1}} \right)/\tau }}} \right)$, where $\tau = 1 / \gamma_\text{rot}$ is the damping time of the rotational motion. Fig.~\ref{fig:gauge}(b) shows the damping time of a levitated particle measured at different pressures \cite{Ahn2020Ultrasensitive}. The damping time is inversely proportional to the pressure. In conclusion, the pressure can be calculated by the corresponding damping rate,
\begin{equation}
	p = \frac{{\pi \rho Rv \gamma_\text{rot}}}{{10 \kappa}}
	\label{eq:pressure}.
\end{equation} 

In high vacuum, the damping time of rotational motion is particularly long but easy to measure. The measured pressure uncertainty in previous experiments is typically within a few percent \cite{Blakemore2020Absolute}. However, in medium vacuum, the damping time becomes much shorter, leading to a large measurement error. Instead, direct measurement of rotation frequency of the levitated particle, which is inversely proportional to the pressure, is a faster method \cite{Ahn2018Optically,Jin2021GHz}. According to earlier experiments, it has been demonstrated that the rotation frequency exhibits a small fluctuation of only 2.3\%, far less than that of ionization gauges (20\% fluctuation). To further enhance the accuracy of the vacuum gauges, cooling the CoM motion of levitated particles can be implemented, resulting in a decreased rotation frequency fluctuation of 0.17\% \cite{Jin2021GHz}. 

In addition to being driven by a circularly polarized laser, a charged particle can also rotate due to the interaction with a spinning electric field generated by two pairs of orthogonal electrodes \cite{Jin2024Quantum,Blakemore2020Absolute}. 
% The rotational motion of charged particles can also be induced by a rotating electric field \cite{Jin2023Quantum,Blakemore2020Absolute}. The rotating electric field can be generated using two pairs of orthogonal electrodes, where 
The four electrodes are applied with four sinusoidal waves with the same frequency and amplitude, but with a $\pi / 2$ phase difference between adjacent electrodes, as shown in Fig.~\ref{fig:gauge}(c) \cite{Blakemore2020Absolute}. The rotation frequency of levitated particles equals to the driving frequency, and satisfies $Id{\omega _r}/dt = {M_\text{ele}} + {M_\text{gas}}$, where the driving torque ($M_\text{ele}$) generated by the electric field ($\left| {\bf{E}} \right|$) is ${M_\text{ele}} = \left| {\bf{p}} \right|\left| {\bf{E}} \right|\sin \varphi$, ${\bf{p}}$ is the dipole moment of the particle, and $\varphi$ is the angle between the electric field and the dipole moment. Consequently, the damping rate of the rotational motion can be given by ${\gamma_\text{rot}} = \left| {\bf{p}} \right|\left| {\bf{E}} \right| \sin \varphi /\left( {I{\omega _r}} \right)$. The angle $\varphi$ can be determined in experiments by analyzing the time domain signals of both the rotational motion of the particle and the driving electric field. Specifically, there is a pressure limit at $\varphi = \pi/2$.

In practical applications, the residual gas species are complicated, such as H$_2$O, He, N$_2$. The vacuum gauge is necessary to be calibrated based on the working environment. 
%For levitated optomechanics, the rotational damping torque on levitated particles is dependent on the residual gas species.
Fortunately, the rotational damping torque on levitated particles depends on the components of residual gases.
Blakemore \textit{et al.} measured the pressure limit of rotation as a function of residual gas species, as shown in Fig.~\ref{fig:gauge}(d) \cite{Blakemore2020Absolute}, indicating the levitation system is even able to detect the gas species.
%On the other hand, using levitated particles can detect the gas species.

\begin{figure*}
	\includegraphics[width=0.8\textwidth]{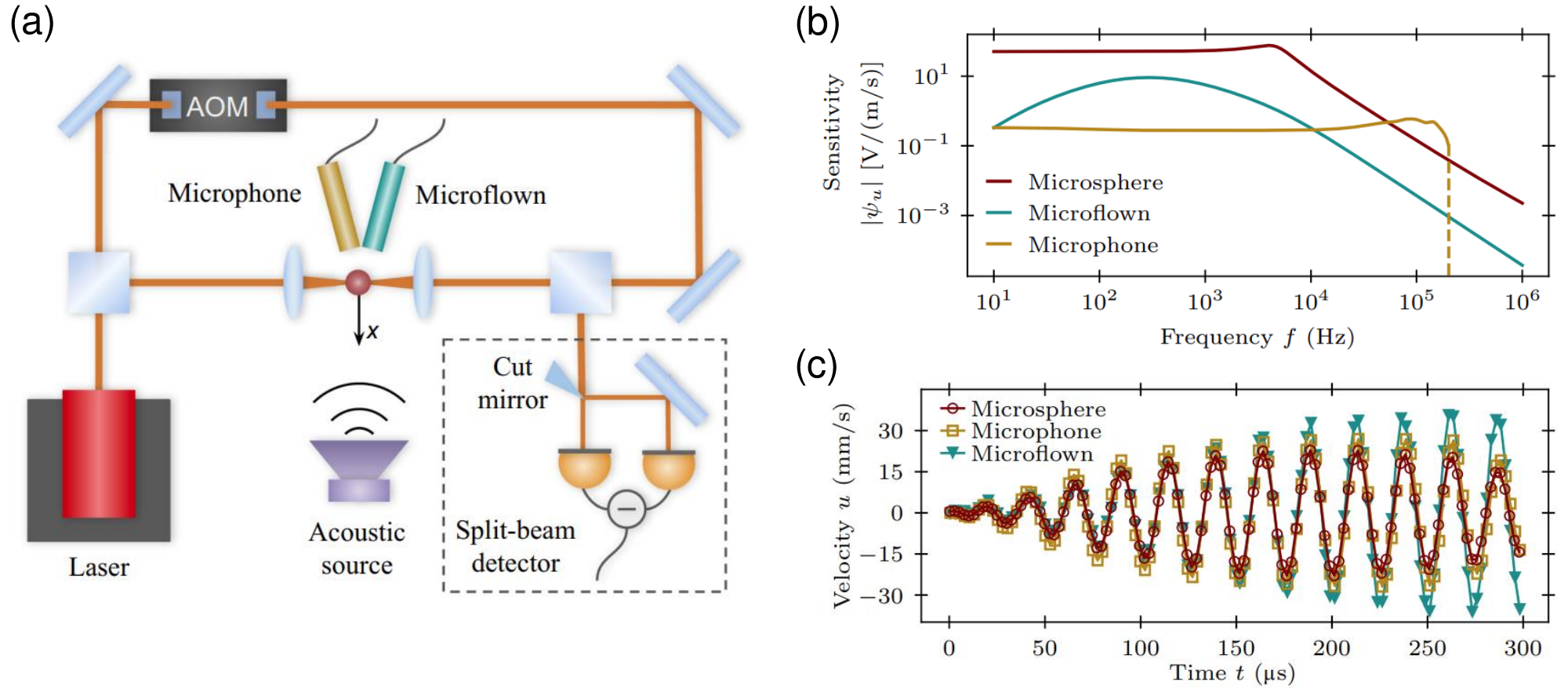}
	\caption{High-bandwidth acoustic transducer, reprinted with permission from ref \cite{Hillberry2024acoustic}, Copyright (2024) by the American Physical Society. (a) Schematic of an acoustic-transducer with a levitated microsphere in a dual-beam optical trap. Two commercial acoustic detectors, a microphone and a Microflown, are placed nearby the trapped particle for comparison. (b) Comparing acoustic detector velocity sensitivities. The levitated microsphere owns a larger bandwidth and performs better at low frequency domain. (c) Tone-burst signal, 10 cycles of 40 kHz, is measured by three acoustic sensors. The levitated microsphere agrees well the commercially calibrated reference sensors.}
	\label{fig:acoustic}
\end{figure*}

Furthermore, the pressure can also be indicated by the CoM motion of levitated particles based on Eq.~\ref{eq:damping} \cite{Dania2023ultrahigh,Barker2023Collision}. Liu \textit{et al.} measured the air pressure around a levitated partile from atmosphere to $5 \times 10^{-6}$ Torr \cite{liu2024nano}. Dania \textit{et al.} measured the damping rate of the CoM motion of a levitated particle at pressures as low as $5 \times 10^{-11}$ Torr, as shown in Fig.~\ref{fig:gauge}(e) \cite{Dania2023ultrahigh}, which is proportional to the pressure. Theoretically, the vacuum gauge based on levitated optomechanics can be used in ultra-high vacuum conditions, even detection of the collision of a signal gas molecule \cite{Yin2011Three,Barker2023Collision}. Yin \textit{et al.} calculated the impact of individual molecule collision on a ground-state cooling particle \cite{Yin2011Three}. The distributions of the particle mean phonon number after a collision with single molecule with two different mass of $m_a = 6.63 \times 10^{-26}$ kg and $m_b = 2.18 \times 10^{-25}$ kg are shown in Fig.~\ref{fig:gauge}(f). 

With rich methods of measuring vacuum pressure, levitated optomechanical system hold the potential to function as highly accurate pressure gauges across a broad pressure range in vacuum environments, offering an advantage over ionization gauges as they are not constrained by residual gas species.

\subsection{High-bandwidth acoustic transducer}
The capability to precisely quantify the motion of levitated particles enables them to function as an acoustic transducer. Recently, Hillberry \textit{et al.} reported a well-performed acoustic sensor with an optically trapped microsphere \cite{Hillberry2024acoustic}. A silica microsphere is levitated in a dual-beam optical trap and two commercial acoustic sensors, a pressure microphone and a Microflown, are placed close to the trapped microsphere for comparison, as depicted in Fig.~\ref{fig:acoustic}(a). After unit calibration, the acoustic detector velocity sensitivities for three sensors are shown in Fig.~\ref{fig:acoustic}(b), indicating the levitated microsphere performs an augmented response to the acoustic velocity signal at low frequency domain and a moderate sensitivity at frequencies exceeding 200 kHz. The levitated microsphere also agrees well with the other two commercial sensors under tone-burst test shown in Fig.~\ref{fig:acoustic}(c). Compared with a microphone or a Microflown, a levitated microsphere can self-calibrate its thermal dynamics without an anechoic room. Since the velocity measured by the microsphere is a vector, this system could be beneficial in sound-source localization.

\begin{figure*}
	\includegraphics[width=0.8\textwidth]{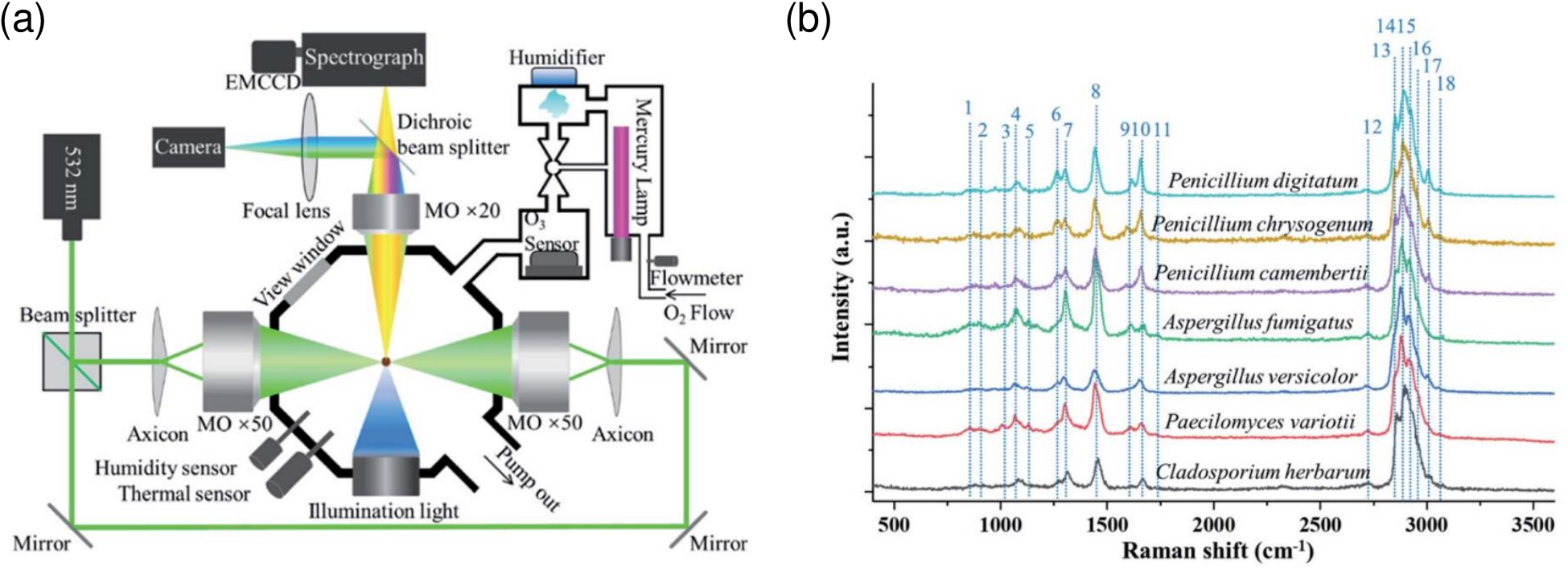}
	\caption{Bioaerosol sensing of levitated optomechanics \cite{Ai2022Characterization}. (a) Schematic of the optical trapping-Raman spectroscopy system. Various single fungus aerosol particles can be levitated in a dual-beam trap at atmospheric pressure without photo-damage and with well controlled environment around the particles. (b) Raman spectra of seven different optically trapped fungus.}
	\label{fig:bioaerosol}
\end{figure*}

\subsection{Chemical and bioaerosol sensing}

In addition to applications in physics and engineering, levitated optomechanics offers potential opportunities in chemistry and biology in atmospheric environment, such as trapping and manipulating individual biomolecules, cells \cite{Souza2010Three} and bioaerosol particles \cite{Ai2022Characterization}. Levitated optomechanics holds great promise for understanding biological systems at the molecular and cellular levels, thereby facilitating biomedical research, and driving innovation in areas such as healthcare and drug discovery. 

The detection of bioaerosol particles in air, i.e., fungi, pollen, bacteria and viruses, has numerous applications across various fields, including public health, environmental monitoring, agriculture, and others. Generally, bioaerosol can be detected by microscope, X-ray spectrometry and Raman spectroscope. However, these methods require performing on substrates or using large amounts of samples, which may change particle properties duo to different environments and contaminants. It is better to perform the study of bioaerosol properties under well-controlled conditions and involve individual airborne particles to avoid particle-surface contact.
% Therefore, studying bioaerosol properties accurately is better under well-controlled environmental conditions and individual particles in air without particle-surface interactions.

By combining the optical levitation and Raman spectroscopy, Ai \textit{et al.} measured the physical, chemical, and biological properties of individual bioaerosol particles under simulated atmospheric conditions \cite{Ai2022Characterization}. Fig.~\ref{fig:bioaerosol}(a) is the schematic of a single bioaerosol particle levitated in a dual-beam trap at atmospheric pressure without photo-damage. Using the technology, they measured Raman spectra of seven different fungus samples in air, as shown in Fig.~\ref{fig:bioaerosol}(b).

\begin{figure*}
	\includegraphics[width=0.8\textwidth]{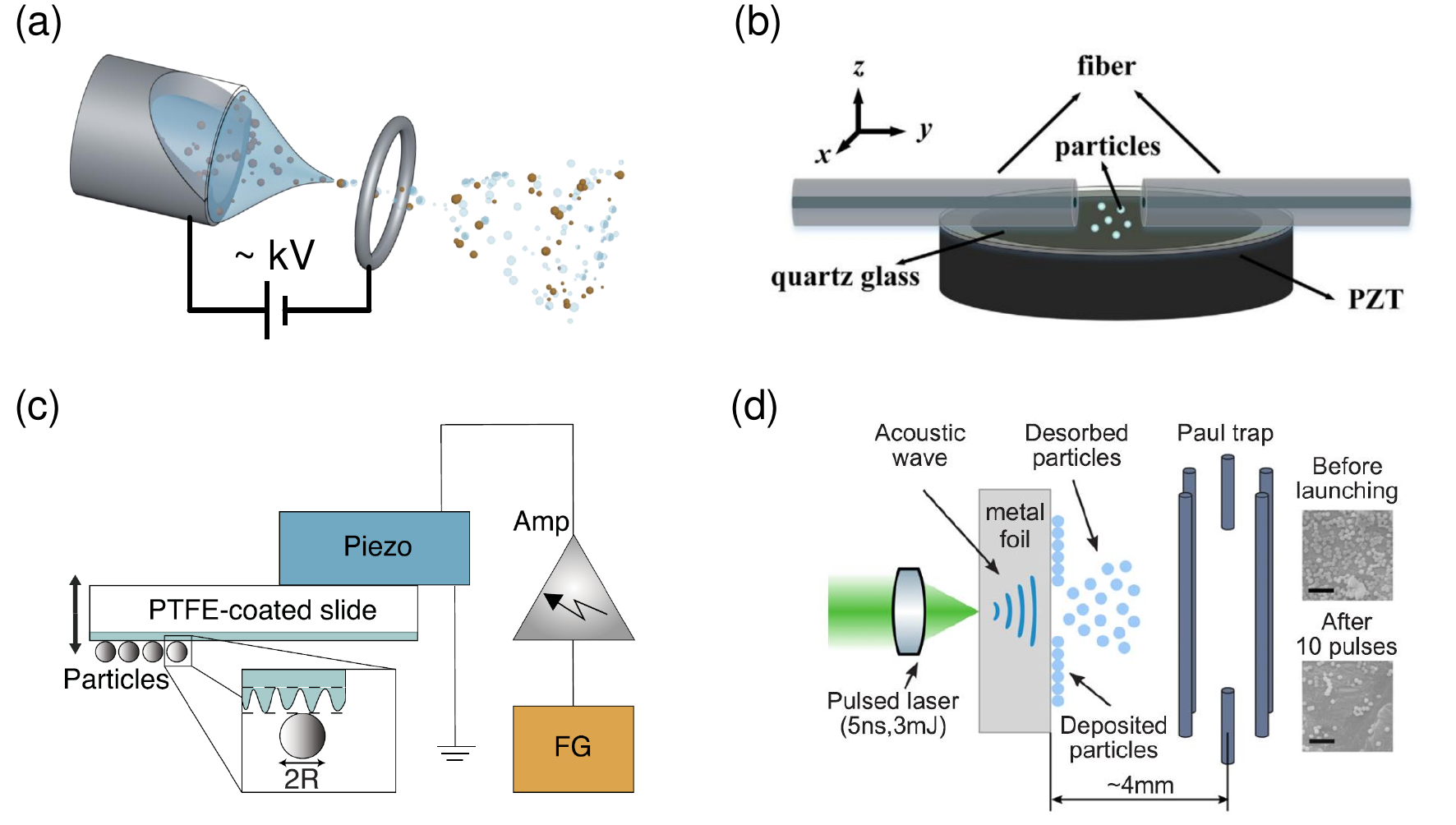}
	\caption{Particle launching techniques. Apart from using an ultrasonic nebulizer to spray particles from a liquid solution, several other methods can be utilized to launch particles into a potential well. (a) Electrospray. The sprayed particles carry a significant number of charges, making them suitable for studying electric forces and levitating in ion traps. (b) and (c) Piezoelectric transducers. (b) is reprinted from ref \cite{Xiao2020A}, Copyright (2020) with permission from Elsevier, and (c) is adapted from \cite{Khodaee2022Dry}. (d) Laser-induced acoustic desorption can be utilized for the direct loading of particles in vacuum, Reprinted from ref \cite{Bykov2019Direct}, with the permission of AIP Publishing.}
	\label{fig:launching}
\end{figure*}

\section{Challenges and opportunities}

\subsection{Particles launching}

As most levitation systems only need to trap a single particle,  an efficient and convenient particle launching approach is necessary. Typically, many experiments use the method that particles are sprayed out from a liquid solution (such as water or isopropanol) with an ultrasonic nebulizer. The particles dispersed in droplets are subsequently guided through a tube and transported to the trapping area. This method is straightforward and applicable at various situations. However, in some special cases, such as experiments involving ion traps \cite{Jin2024Quantum,Conangla2018Motion}, the charge-to-mass ratio of particles may be insufficient for stable levitation. The number of charges carried on particles is limited to a few \cite{Frimmer2017Controlling,Ju2023Near}. In hybrid traps with an optical cavity \cite{Delic2020Cooling,Millen2015Cavity}, excessive particles and solution droplets sprayed into the chamber may adhere to the surface of the optical cavity, reducing its quality factor. Additionally, this method requires opening the vacuum chamber every time to load particles, which is harmful to high vacuum experiments. Recently, sublimation-activated release (SAR) loading technique using the sublimation of camphor was used to selectively load microparticles into a magneto-gravitational trap \cite{murphy2024selective}.  

To increase the charge-to-mass ratio of particles, electrospray can be employed. The schematic of an electrospray is shown in Fig.~\ref{fig:launching}(a). Particles in liquid solution (usually alcohol or isopropanol) is pumped through a capillary metal tube. A DC high voltage of a few kV is applied on the metal tube, leading to an increasing charge accumulation on a liquid droplet. Once the repulsive electric force is larger than its surface tension, the droplet is divided into smaller droplets, each containing a single particle. Subsequently, the particles are accelerated and sprayed out by the electric field formed between the metal tube and the grounded electrode. Typically, a particle with a diameter of 1 $\mu$m prepared via electrospray carries a charge number in between 1,000 and 10,000 \cite{Jin2024Quantum}, which is large enough for stable levitation in an ion trap.

As we mentioned, the traditional particle loading by ultrasonic nebulizer is inefficient and may contaminate the vacuum chamber, which wastes amount of time to re-establish the vacuum condition. Direct particle loading in a vacuum environment provides a viable solution to this issue. One approach involves the use of piezoelectric transducers \cite{Li2013Fundamental,Xiao2020A,Khodaee2022Dry}, as shown in Figs.~\ref{fig:launching}(b) and (c). Initially, particles are deposited onto a glass substrate affixed to a piezo element. Large particles, with radii on the order of micrometers, can be easily ejected from the glass substrate. However, when dealing with small nanoparticles, the primary challenge lies in overcoming the strong attractive force between the particles and the glass substrate. This obstacle can be mitigated by using a polytetrafluoroethylene (PTFE) coated substrate, which reduces the adhesive forces (Fig.~\ref{fig:launching}(c)), which has been demonstrated to successfully launch a nanoparticle with a radius of 43 nm \cite{Khodaee2022Dry}.

An alternative approach for direct particle loading in vacuum involves the impact force of a pulsed laser to lift the particles from a substrate, which is named Laser-Induced Acoustic Desorption (LIAD) (Fig.~\ref{fig:launching}(d)) \cite{asenbaum2013cavity,Bykov2019Direct}. Initially, a liquid solution containing particles is dried on a 250 $\mu$m thick aluminum foil. A focused pulse laser from the backside of the foil generates an acoustic shock wave that launches the particles from the front side through a process known as acoustic desorption. To enhance loading efficiency, a four-rod Paul trap with a broad trapping range and a deep potential well is positioned close to the foil. The launched particles are first trapped in the Paul trap and can subsequently be delivered to an optical trap. This approach proves particularly advantageous for loading particles in ultra-high vacuum environments. The ability to directly load particles into a Paul ion trap at pressures down to  $10^-7$ Torr has been experimentally demonstrated \cite{Bykov2019Direct}.

\subsection{On-chip levitation}

\begin{figure}
	\includegraphics[width=0.45\textwidth]{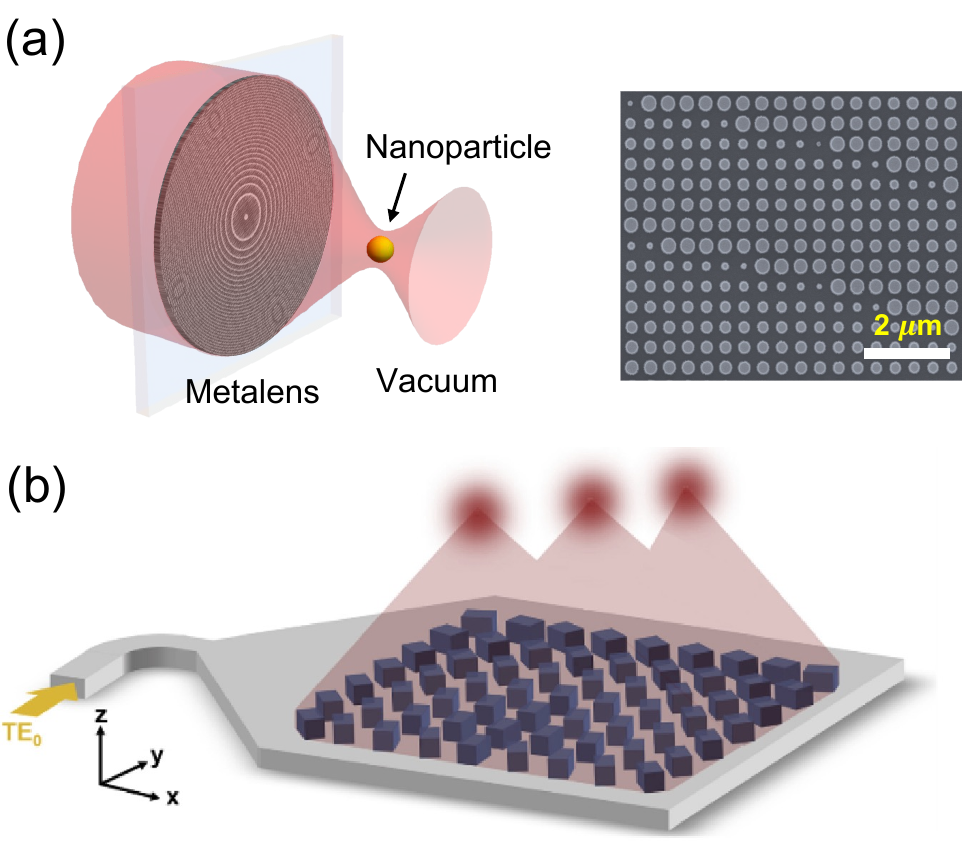}
	\caption{On-chip trapping and detection of levitated optomechanics. (a) Schematic of an optically levitated particle using a metalens achieved by Shen at al. \cite{Shen2021Onchip}. The inset is the SEM image of the metalens, consisting of silicon nanopillars with a height of 500 nm. (b)
	Schematic of a wave-driven metalens for generating multi-trap emission. Nanopillars are arranged in a square lattice. Adapted with permission from ref \cite{Yu2024integrated} \copyright Optical Society of America.
	}
	\label{fig:onchip}
\end{figure}

Levitated optomechanics is widely used for a range of  applications, as discussed in the previous sections. Minimization and integration are crucial considerations in the practical deployment of levitated devices. On-chip levitation emerges as a promising platform to achieve compact levitation systems.
% In the typical experiments using high NA objective lenses, the optical trap and detection systems involve complex and sizable light paths. Both of these components can be replaced with smaller elements to achieve a more compact package.

The high NA lens used for laser focusing and signal collection can be replaced by a metalens, as demonstrated by Shen \textit{et al.} \cite{Shen2021Onchip} and shown in Fig.~\ref{fig:onchip}(a). The metalens is constructed by arranging phase-shifting elements on a surface to create a phase profile analogous to that of a traditional lens. It is designed with a diameter of 425 $\mu$m and a focal length of 100 $\mu$m. It has an NA of 0.9 for a 1064 nm laser in vacuum.
With this thin and compact metalens, a silica nanoparticle could be levitated in vacuum with a trapping laser power of 200 mW.

\begin{figure*}
	\includegraphics[width=1\textwidth]{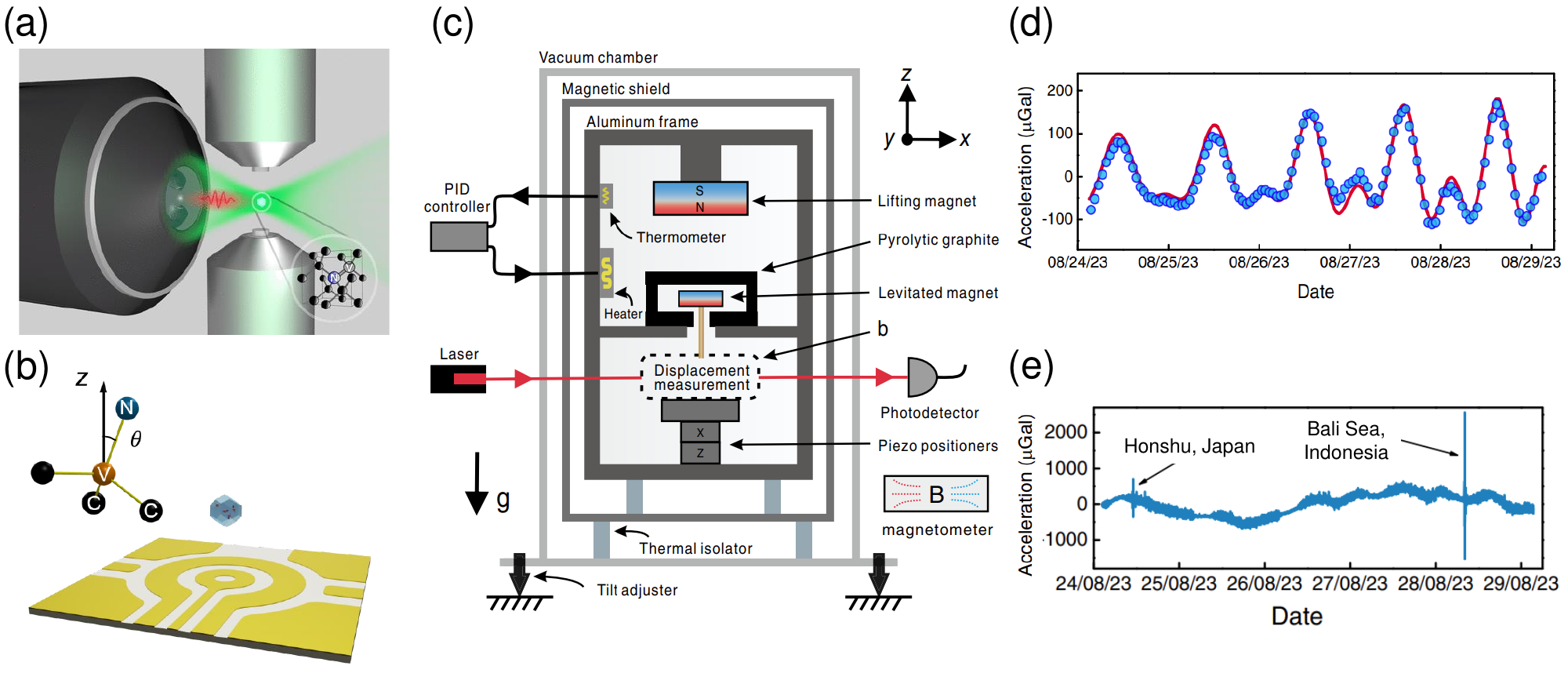}
	\caption{Ion trap and magnetic trap. (a) Schematic of an ion trap composed of two tip electrodes used to trap and control charged diamond particles. Adapted with permission from ref \cite{Conangla2018Motion}. Copyright (2018) American Chemical Society. (b) Schematic of a surface ion trap \cite{Jin2024Quantum}. The $\Omega$-shaped circuit is applied AC high voltage and the central ring electrode is grounded to levitate particle above surface. Combining a microwave on the $\Omega$-shaped circuit can facilitate control of the embedded NV centers for diamond particles. 
		%(c) Potential simulation in the $xz$-plane of a magnetic trap consisted of two pole pieces passively driven by a NdFeB permanent magnet \cite{Houlton2018Axisymmetric}. The magnetic field generated combined with the gravity forms a three-dimensional confining potential for levitating diamagnetic particles. (d) Optical image of a magnetically levitated particle. (e) Levitation of a pyrolytic graphite disk using magnetic stage consisted of NdFeB magnets and the arbitrary control of the moving direction of the graphite disk by a laser \cite{Kobayashi2012Optical}. (f) Images of the linear motion of the pyrolytic graphite disk in one direction.
		(c) Schematic depiction of the diamagnetic levitation gravimeter. A micro-oscillator with a mass of 215 mg is levitated in the shield of pyrolytic graphite. (d) Comparison with theoretical data of the earth tides. The recorded acceleration data with drifting removed is represented as blue dots while the red line is the theoretical data. The experimental data agrees well with theoretical calculation by a high correlation coefficient as high as 0.97. (e) Raw experimental data displays two sharp signals, indicating the effects of two seismic events nearby Honshu, Japan and Bali Sea, Indonesia, respectively. (c), (d) and (e) are reprinted with permission from ref \cite{Leng2024EarthTide}, Copyright (2024) by the American Physical Society.
	}
	\label{fig:traps}
\end{figure*}

In contrast to a conventional objective lens, the metalens is superior at operating under more extreme conditions and provides greater flexibility in generating complicated trapping potentials through nanofabrication technique \cite{Yu2024integrated,sun2024tunable}. Recently, Yu \textit{et al.} \cite{Yu2024integrated} numerically demonstrated a wave-driven metalens for creating optical tweezer arrays. The nanopillars, made of silicon, are arranged in a square lattice as shown in Fig.~\ref{fig:onchip}(b). The incident light at the taper with transversal electric mode ($\text{TE}_0$) couples with the scattered wave and deliver extra phase, resulting in a multi-trap phase profiles. This compact integrated design of optical metalenses get rid of bulk spatial light modulators necessitated by conventional optical tweezer arrays.

Furthermore, a pair of optical fibers, carrying two linearly polarized counter-propagating lasers to form a standing wave, can be used for particle levitation \cite{Xiao2020A,Melo2023Vacuum}. 
The nanoparticles are confined within the stationary wave pattern formed by the interference of two counter-propagating beams. Concurrently, the motion signals of the levitated particles can be detected using another pair of orthogonal fibers \cite{Kuhn2017Full,Melo2023Vacuum}, offering an advantage in better adapting to the scattering pattern of the particle. Additionally, a bottom layer of the hybrid on-chip trap with planar electrodes can be designed for cooling the CoM motion of levitated particles \cite{Jin2024Quantum,Melo2023Vacuum}.

\subsection{Optomechanics with ion traps and magnetic traps}

Optical levitation systems demonstrate high trapping frequencies (> 100 kHz) for nanoparticles, regardless of the charge and magnetic susceptibility of levitated objects. The motion of optically levitated particles is easily controlled and measured. However, laser recoil heating and particle absorption limit the stable levitation in high vacuum. Previous optical levitation experiments have exclusively used silica particles in high vacuum for precision measurements due to their low optical absorption and high-temperature thermal stability. However, optically levitating other materials is constrained to low vacuum conditions as detailed in Table~\ref{tbl:particles}. As the dimension of the levitated particle increases to the micrometer scale, the trapping frequency rapidly drops to tens of hertz. To address these limitations, ion traps and magnetic traps provide viable alternatives, enabling particle levitation in high vacuum with negligible heating effects.

Ion traps create a three-dimensional potential for charged particles by employing a combination of static and AC quadrupole potentials. Various designs of ion traps have been reported, including the ring trap \cite{Delord2020Spin}, four-rod trap \cite{Bykov2019Direct,Dania2022Position}, a pair of tips trap \cite{Millen2015Cavity,Conangla2018Motion}, and surface trap \cite{Wang2015Surface,Jin2024Quantum}. 
% The three-dimensional ion traps exhibit a harmonic potential in the central region. 
The stable levitation of micro- and nano-scale particles, such as diamonds, with an ion trap has been demonstrated \cite{Conangla2018Motion,Delord2020Spin,Jin2024Quantum}. Conangla \textit{et al.} used an ion trap formed with a pair of tip electrodes to levitate charged diamonds, as shown in Fig.~\ref{fig:traps}(a). 
% The stably levitated diamonds in high vacuum was first achieved by Jin \textit{et al.} using a surface ion trap \cite{Jin2023Quantum}. 
The  optical readout of electron spins in a levitated diamond in high vacuum with a surface ion trap was first demonstrated by Jin \textit{et al.} \cite{Jin2024Quantum}. The surface ion trap is created on a sapphire substrate with a crucial part of only about 2 mm $\times$ 2 mm $\times$ 0.4 mm in size, as shown in Fig.~\ref{fig:traps}(b). The levitated nanodiamond's internal temperature remains stable at about 350 K under a pressure of $6 \times 10^{-6}$ Torr. The temperature is still moderate for quantum control of the NV spins embedded in the diamonds. In addition, the levitated diamonds can be driven to rotate at a frequency of up to 20 MHz, surpassing typical NV center electron spin dephasing rates. 
% Barrnet effect and Berry phase arising from particle rotation can be study. 
Delord \textit{et al.} demonstrated the strong spin-mechanical coupling of levitated diamonds in an ion trap \cite{Delord2020Spin}, offering the potential for using spins to create non-classical states of motion.

Magnetic levitation, including diamagnetic levitation \cite{Simon2000Diamagnetic} and diamagnetically stabilized magnet levitation \cite{Geim1999Magnet}, is an alternative technique for object levitation. The method is widely used in extensive applications, including force and acceleration sensors \cite{Ando2018A}, fluid viscosity sensors \cite{Clara2015A} and gas flowmeters \cite{Zhang2018Design}. Considering a diamagnetic particle levitated in an external magnetic field (\textbf{B}), the potential energy of the particle can be expressed as $U = - \chi {B^2} V / \left(2 \mu _0\right) + mgy$ \cite{Simon2000Diamagnetic}, where the two terms are attributed to the external magnetic field and the Earth's gravity. $\chi$, $V$, and $m$ are the magnetic susceptibility, volume, and mass of the levitated particle, respectively. $\mu_0$ is vacuum permeability, $g$ is gravitational acceleration, and $y$ is the position in the vertical direction. The stable trapping condition for diamagnetic particles depends on the equilibrium between the magnetic force acting on the particle and gravity. 

Recently, Leng \textit{et al.} reported a low-drift room-temperature gravimeter based on diamagnetically stabilized magnet levitation with an acceleration sensitivity of 15 $\mu$Gal/$\sqrt{\text{Hz}}$ and a drift of 61 $\mu$Gal/$\sqrt{\text{Hz}}$ per day (the best among relative gravimeters) \cite{Leng2024EarthTide}. The levitated micro-resonator with a proof mass of 215 mg is located in the aluminum frame with a magnetic shield covered in vacuum chamber, shown in Fig.~\ref{fig:traps}(c). Fig.~\ref{fig:traps}(d) illustrates the recorded experimental acceleration data presented alongside theoretically calculated data of earth tides for comparison, showing a high correlation coefficient of 0.97. The earthquake events can also be recorded and manifested as spikes in the raw data, shown in Fig.~\ref{fig:traps}(e).

\section{Conclusion}

Levitated optomechanics, an area that has rapidly progressed since 2010 \cite{Li2010Measurement,Romero-Isart_2010,chang2010cavity}, holds significant promise for diverse applications owing to its high force, acceleration, and torque sensitivities, which have been demonstrated in laboratory environments \cite{Ranjit2016Zeptonewton,Monteiro2020Force,Liang2023Yoctonewton,Ahn2020Ultrasensitive}. Levitated objects are also sensitive to external fields and variations, rendering them attractive candidates for high-precision sensors like accelerometers or gyroscopes. A compact accelerometer  with an optically levitated particle has been installed on a vehicle running on a road \cite{Han2023Feedback}, showing its potential for real-world applications. Levitation systems are suitable for manipulation in vacuum and microgravity environments, showcasing potential advancements in instruments for future space missions.
Levitation systems also offer the ability to conduct non-destructive testing and characterize materials at the nanoscale. 
While these applications hold promise, it is necessary to acknowledge challenges in implementing levitation systems beyond the laboratory environments. Addressing technical issues such as particle loading in vacuum, system minimization, stability maintenance, and scalability for practical applications remains a crucial focus in this evolving field.

\begin{acknowledgments}
We thank the support from the Office of Naval Research under Grant No. N00014-18-1-2371, the National Science Foundation under Grant PHY-2110591, and the Gordon and Betty Moore Foundation. 
\end{acknowledgments}

%\bibliography{references}


\end{document}